\documentclass[pra,twocolumn,superscriptaddress,showpacs,amsmath,amstex,amssymb,citeautoscript]{revtex4-1}

\newcommand{\vect}[1]{\boldsymbol{#1}}

\usepackage[T1]{fontenc}
\usepackage[latin9]{inputenc}
\usepackage{lipsum}
\usepackage{amsmath}
\usepackage{amssymb}
\usepackage{bbm}
\usepackage{braket}
\usepackage{xcolor}
\usepackage{pifont}
\usepackage[mathscr]{euscript}
\usepackage[shortlabels]{enumitem}

\allowdisplaybreaks

\usepackage{graphicx}
\usepackage{dsfont}
\usepackage[colorlinks=true]{hyperref}  
\hypersetup{
    bookmarks=true,         
    unicode=false,          
    pdftoolbar=true,        
    pdfmenubar=true,        
    pdffitwindow=false,     
    pdfstartview={FitH},    
    pdftitle={Microscopic theory of superconductivity in twisted double-bilayer graphene},    
    pdfauthor={Samajdar and Scheurer},     
    pdfsubject={},   
    pdfcreator={},   
    pdfproducer={}, 
    pdfkeywords={} {} {}, 
    pdfnewwindow=true,      
    colorlinks=true,       
    linkcolor=blue, 
    citecolor=blue,        
    filecolor=magenta,      
    urlcolor=blue           
}

\newcommand{\equref}[1]{Eq.~(\ref{#1})}

\newcommand{\figref}[1]{Fig.~\ref{#1}}
\newcommand{\refcite}[1]{Ref.~\onlinecite{#1}}

\newcommand{\appref}[1]{Appendix~\ref{#1}}
\newcommand{\mc}{\mathcal}
\newcommand{\coI}{\text{intra}}
\newcommand{\coII}{\text{inter}}

\newcommand{\pdagger}{{\phantom{\dagger}}}
\newcommand{\diff}{\mathrm{d}}

\renewcommand{\approx}{\simeq}

\definecolor{wrongultramarine}{rgb}{1,0.5,0}

\begin{document}

\title{Microscopic pairing mechanism, order parameter, and disorder sensitivity in moir\'e superlattices: Applications to twisted double-bilayer graphene}

\author{Rhine Samajdar}
\affiliation{Department of Physics, Harvard University, Cambridge, MA 02138, USA}
\author{Mathias S. Scheurer}
\affiliation{Department of Physics, Harvard University, Cambridge, MA 02138, USA}

\begin{abstract}
Starting from a continuum-model description, we develop a microscopic weak-coupling theory for superconductivity in twisted double-bilayer graphene. We study both electron-phonon and entirely electronic pairing mechanisms. 
In each case, the leading superconducting instability transforms under the trivial representation, $A$, of the point group $C_3$ of the system, while the subleading pairing phases belong to the $E$ channel. We explicitly compute the momentum dependence of the associated order parameters and find that the leading state has no nodal points for electron-phonon pairing but exhibits six sign changes on the Fermi surface if the Coulomb interaction dominates. 
On top of these system-specific considerations, we also present general results relevant to other correlated graphene-based moir\'e superlattice systems. We show that, irrespective of microscopic details, triplet pairing will be stabilized if the collective electronic fluctuations breaking the enhanced $\text{SU}(2)_+ \times \text{SU}(2)_-$ spin symmetry of these systems are odd under time reversal, even when the main 
$\text{SU}(2)_+ \times \text{SU}(2)_-$-symmetric part of the
pairing glue is provided by phonons. Furthermore, we discuss the disorder sensitivity of the candidate pairing states and demonstrate that the triplet phase is protected against disorder on the moir\'e scale.
\end{abstract}

\maketitle

\section{Introduction}

The family of moir\'e superlattice systems displaying correlated physics has been expanding rapidly \cite{macdonald2019bilayer}. One such heterostructure that has generated much interest is twisted double-bilayer graphene (TDBG) \cite{2019arXiv190306952S,ExperimentKim,PabllosExperiment,burg2019correlated}, in which two AB-stacked graphene bilayers are twisted relative to each other. This system stands out not only due to the additional tunability of the band structure via electric fields \cite{2019PhRvB..99g5127Z,2019arXiv190108420R,2019arXiv190300852C,2019arXiv190308685L,koshino2019band,FirstModel,2019arXiv190600623H}, but also because both the gap of the correlated insulating state at half-filling and the critical temperature of the superconducting phase can be enhanced by application of a magnetic field, which hints that its physics is different from that of twisted bilayer graphene (TBG) \cite{2018Natur.556...80C,SuperconductivityTBG,Yankowitz1059,2019arXiv190306513L}. The superconducting phase of TDBG is found to be very fragile \cite{PabllosExperiment,burg2019correlated}: it only emerges \cite{2019arXiv190306952S,ExperimentKim} on the electron-doped side relative to the insulator at half-filling of the conduction band and quickly weakens away from this optimal doping value.

In a recent publication \cite{2019arXiv190603258S}, we performed a systematic classification of the possible pairing instabilities in TDBG and other related moir\'e superlattice systems, using general constraints based on symmetries and energetics. Here, we complement \refcite{2019arXiv190603258S} with an explicit microscopic calculation using a continuum model for the underlying graphene sheets, inspired by 
theoretical studies of superconductivity in TBG \cite{fidrysiak2018unconventional, PhysRevB.98.195101, 2018PhRvB..98v0504P,2018PhRvL.121y7001W, 2018PhRvB..98x1412C,lian2019twisted, alidoust2019symmetry, wu2019topological, YiZhuangPairing, hu2019geometric, huang2019antiferromagnetically,julku2019superfluid,2019arXiv190903514W}. 
While there have been indications pointing towards an electron-phonon based pairing mechanism in TBG \cite{2019arXiv191109198S,2019arXiv191113302S}, recent experiments \cite{2020arXiv200311072L} seem to challenge these conclusions. In any case, the situation as regards TDBG is, at present, completely open. 
As a first step towards understanding the pairing glue in TDBG, in this work, we analyze both electron-phonon (conventional) and electron-electron (unconventional) interactions-based  pairing. The symmetry properties and gap structures of the leading and subleading instabilities, and their doping dependencies are computed and discussed in relation to experiment. We also describe the possibility that the electron-phonon coupling provides the main pairing glue but the dominant interactions that distinguish between singlet and triplet stem from purely electronic processes. Additionally, we examine the disorder sensitivity of the different pairing states. 

This article is organized as follows. As a starting point, in Sec.~\ref{sec:model}, we introduce the continuum model for TDBG that serves as the basis for our microscopic computations. Next, in Secs.~\ref{sec:eph} and \ref{sec:ee}, we examine pairing mechanisms based on electron-phonon and electron-electron interactions, respectively, and evaluate the basis functions for the associated superconducting states. At the end of Sec.~\ref{sec:ee}, we also discuss the interplay of the two pairing mechanisms. Proceeding further, we also consider the impact of disorder on these states in Sec.~\ref{sec:disorder}. Finally, four appendices, \ref{app:model} through \ref{app:singlettriplet}, supplement our discussion in the aforementioned sections with the details of our calculations.

\section{Model and band structure}
\label{sec:model}

\begin{figure}[tb]
\includegraphics[width=\linewidth]{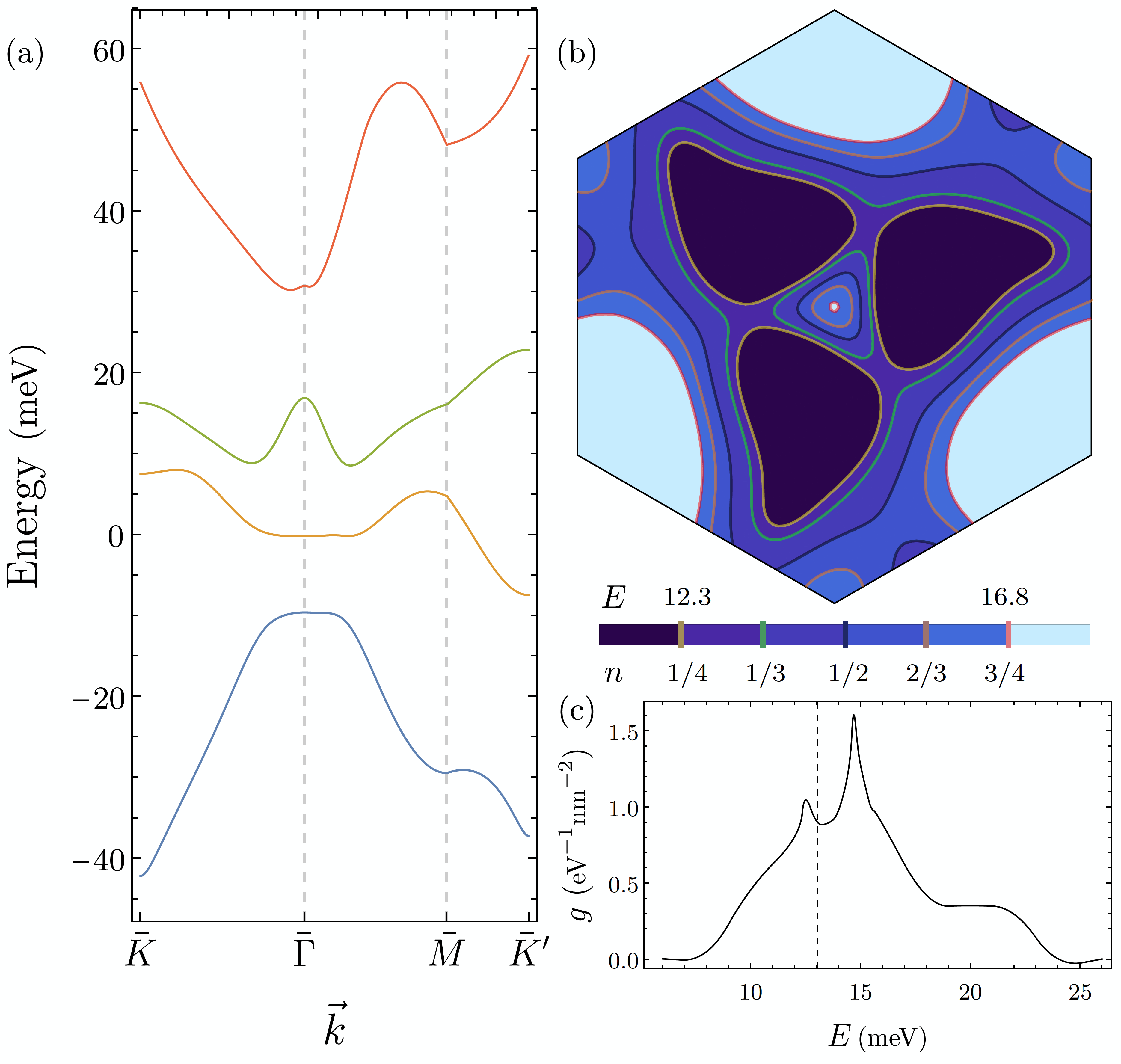}
\caption{\label{fig:band}(a) Band structure of TDBG for the $\xi$\,$=$\,$+$ valley at a twist angle $\theta$\,$=$\,$1.24^\circ$ and a displacement field (the electrostatic energy difference between adjacent sheets) of $d$\,$=$\,$5$ meV; the latter can be varied to controllably tune the dispersion. The band hosting superconductivity is indicated in green.\,(b) Fermi surfaces corresponding to five different fillings of this band. (c) The density of states per spin per valley as a function of energy.}
\end{figure}

The TDBG heterostructure is composed of two individually aligned Bernal-stacked bilayer graphene (BLG) sheets, twisted relative to each other by a small angle $\theta \ll 1$, forming a moir\'e superlattice. When the two BLGs are decoupled, the spectrum of each exhibits a familiar pair of linearly dispersing Dirac cones \cite{morell2010flat,PhysRevB.85.195458}; we denote the location of the Dirac points of the $l$-th BLG by $\vect{K}^l_\xi$, where $\xi$\,$=$\,$\pm$ labels the two valleys. Upon twisting, one moves away from this limit and the superlattice structure now manifests itself in the coupling between the lower layer of the top BLG and the topmost sheet of the bottom stack. This hybridizes the eigenstates at a Bloch vector $\vect{k}$ in the moir\'{e} Brillouin zone (MBZ) with those 
related by reciprocal lattice vectors of the moir\'{e} superlattice.

The resulting electronic band structure of TDBG is well described by a continuum model \cite{koshino2019band, dos2007graphene,bistritzer2011moire,dos2012continuum,weckbecker2016low}, reviewed in Appendix~\ref{app:model}. The key simplification achieved by such a continuum model vis-\`{a}-vis microscopic lattice descriptions is that it restores periodicity \cite{balents2019general}, facilitating the application of Bloch's theorem to a system that, for generic twist angles, will only be quasiperiodic. Our low-energy Hamiltonian [Eq.~\eqref{eq_AB-AB}] is comprised of block-diagonal elements, which are $4\times4$ matrices describing the Bloch waves of individual graphene bilayers at different momenta \cite{PhysRevB.99.205134, PhysRevResearch.1.013001}. The interlayer coupling breaks the translation invariance of the (AB-stacked) BLG unit cell and couples these blocks to one another. Nevertheless, since these terms still have the translational symmetry of the moir\'{e} supercell, it is convenient to transform the Hamiltonian from real space to a momentum basis. For $\vect{k}$ in the MBZ, the eigenenergies $E_{\vect{k},\xi}$ can be computed independently for each $\xi$ as the intervalley coupling is negligible for small twist angles.

For an accurate description of the band structure, we incorporate lattice relaxation effects \cite{nam2017lattice, koshino2018maximally}, which are known to be important for isolating flat bands. 
Moreover, on including terms accounting for trigonal warping \cite{jung2014accurate} and particle-hole asymmetry \cite{mccann2006landau}, the first conduction and valence bands, around charge neutrality, always acquire a sizable dispersion and unlike TBG \cite{morell2010flat, tarnopolsky2019origin, Khalaf_2019}, there no longer exists a sharp ``magic angle'' at which they are almost perfectly flat. For small $\theta$, these bands overlap with each other but they can be separated by a gate voltage applied between the top and bottom layers, or, in other words, a displacement field.  
The $10$--$15$~meV bandwidth of the band hosting superconductivity, shown in green in \figref{fig:band}(a), is still smaller than the interaction scale \cite{2019arXiv190308685L}, revealing the system's strongly correlated nature. Adding to the distinction from TBG is the fact that the physics of TDBG is dominated by a \textit{single} band (per spin per valley)---rather than two---as a consequence of the broken twofold rotational symmetry $C_2$ \cite{2019arXiv190308685L}, which formerly protected the Dirac points in TBG.
Bearing in mind the experimental results, in the following, we focus on this narrow band for a specific twist angle, $\theta$\,$=$\,$1.24^\circ$, at which superconductivity was detected in TDBG samples \cite{ExperimentKim}. The associated Fermi surfaces for different fractional fillings of the relevant band, and the density of states (DOS) are sketched in Figs.~\ref{fig:band}(b) and (c). As is often the case for graphene-based moir\'{e} systems \cite{koshino2018maximally, PasupathySTM}, we note the existence of van Hove singularities in the noninteracting DOS at the energies of the saddle points in the band structure \cite{hsu2020topological, 2019arXiv190607302W}.

\begin{figure*}[htb]
\includegraphics[width=\linewidth]{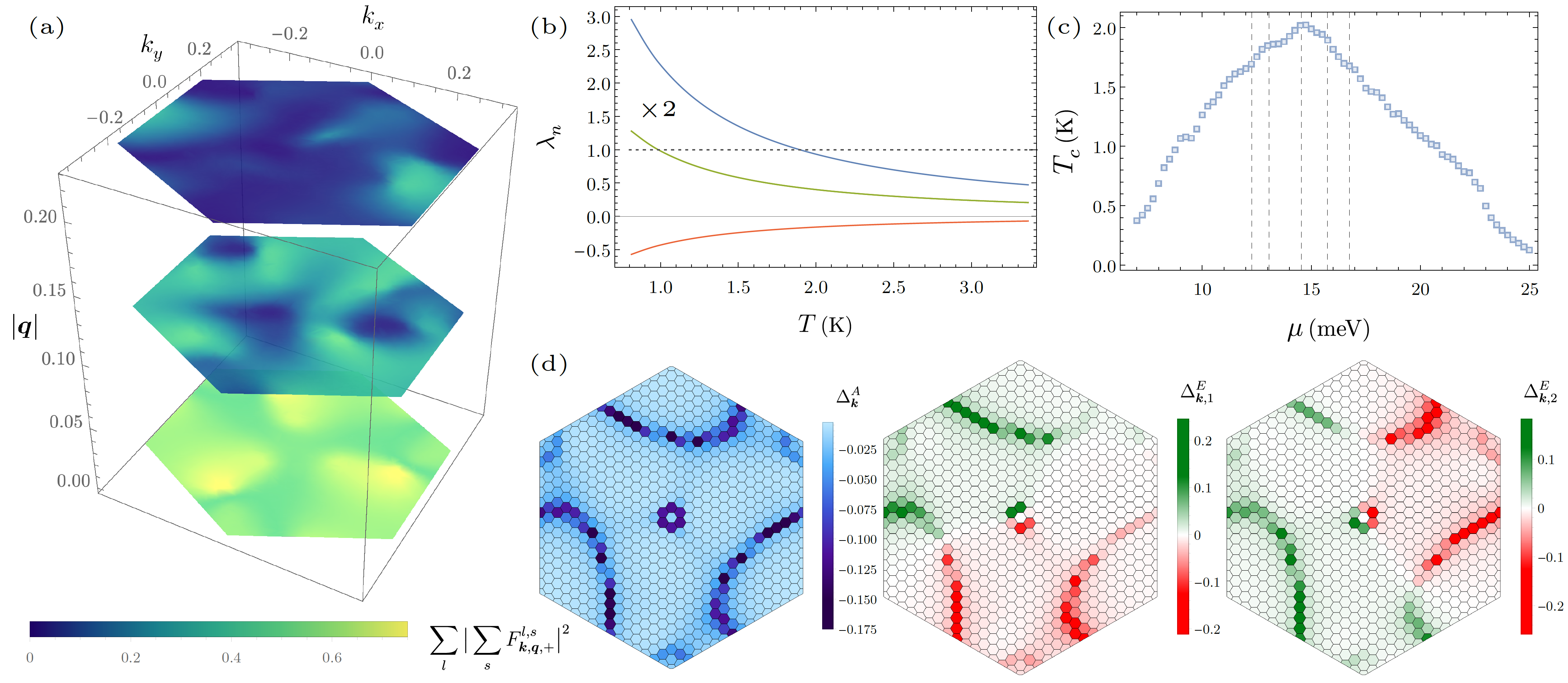}
\caption{\label{fig:Phonon}Phonon-mediated superconductivity in TDBG. (a) The form factors associated with the projection of the densities on to the lowest band. In the $\vect{q}$\,$=$\,$0$ plane, the threefold rotational symmetry of the lattice is directly visible. (b) The four largest eigenvalues $\lambda$ of the matrix $\mathcal{M}$ as a function of temperature $T$; the subleading eigenvalues are doubly degenerate, as marked by $\times 2$. (c) Dependence of the superconducting transition temperature on the chemical potential; dashed lines indicate the five integer fillings of Fig.~\ref{fig:band}. (d) Basis functions of the leading (left) and subleading (middle and right) superconducting states at $2/3$ filling. Note that the extrema of the order parameter trace out the Fermi surface in \figref{fig:band}(b). 
}
\end{figure*}

\section{Conventional pairing}
\label{sec:eph}
To begin our analysis of superconductivity, we first consider the option that the attractive interaction between electrons is mediated by the phonon modes of the two-dimensional graphene layers \cite{2019arXiv190607302W}. This is a natural extension of the physics of magic-angle TBG in which the phonon-driven electron-electron attraction, in combination with an enhanced DOS $\sim\,$ $10$ eV$^{-1}\,$nm$^{-2}$ [nearly ten times larger than $g\,(E)$ in Fig.~\ref{fig:band}], has been predicted to induce intervalley pairing \cite{2018PhRvL.121y7001W, lian2019twisted, 2018PhRvB..98x1412C, 2018PhRvB..98x1412C, wu2019phonon}, yielding a critical temperature $T_c$\,$\sim$\,$1$~K \cite{SuperconductivityTBG}. Given that a similar enhancement of the DOS is lacking in TDBG due to the significantly larger bandwidth, one may question whether such a mechanism still holds in this system.

\subsection{Electron-phonon interactions}\label{ElectronPhononCoupl}

To this end, we consider, in particular, acoustic phonon modes, prompted by previous works on TBG \cite{lian2019twisted,wu2019phonon}. While the layer-symmetric modes (all four layers moving together) and the modes where the two BLGs move against each other (corresponding to a shift of the moir\'e lattice) remain gapless, the phonons that shift any of the sheets of the same BLG against each other acquire a mass. Concentrating on the former modes, we assume that their velocity is the same, motivated by the expected weak interlayer phonon coupling \cite{yan2008phonon}. The electron-phonon Hamiltonian \cite{wu2019phonon} can, therefore, be written as a coupling of acoustic phonon modes, $\alpha^{}_{\vect{q},l}$, and the electron density $\hat{\rho}_{\vect{q},l}$ of the lower ($l$\,$=$\,$1$) and upper ($l$\,$=$\,$2$) BLG,
\begin{alignat*}{1}
H^{}_{\mathrm{EP}}= \frac{D}{\sqrt{N\Omega}} \sum_{\vect{q},l} \sqrt{\frac{\hbar}{2\, \bar{m}\, \omega_{\vect{q}}}} \left(-i \vect{q}.\hat{e}_{\vect{q}} \right) \left(\alpha^{\pdagger}_{\vect{q},l} + \alpha^\dagger_{-\vect{q},l} \right)\hat{\rho}^{}_{\vect{q},l}.
\end{alignat*}
Here, $\bar{m}$ is the mass density, $N$ and $\Omega$ denote the number and area of moir\'e unit cells, respectively, $\hat{e}_{\vect{q}}$ is the displacement unit vector, and $\omega_{\vect{q}}$\,$=$\,$v_{ph} \lvert \vect{q} \rvert$ the phonon frequency; we take $v_{ph}$\,$=$\,$2$\,$\times$\,$10^6\,$cm/s \cite{wu2019phonon}. The deformation potential $D$\,($\sim$\,$25\,$eV) is much larger than the interlayer tunneling parameter ($\sim$\,$0.1\,$eV) \cite{wu2019phonon}, which allows for neglecting the effects of the latter. The Debye frequency of moir\'{e} phonon bands \cite{lian2019twisted} is $\sim$\,$v_{ph}\,\lvert q_\textsc{m} \rvert/\sqrt{3}$\,$=$\,$4.85\,$meV, defining $q_\textsc{m}$\,$\equiv$\,$\lvert\vect{G}^{\rm M}_{1,2}\rvert$ with primitive vectors $\vect{G}^{\rm M}_{1,2}$ of the reciprocal moir\'e lattice. Thus, in contrast to TBG \cite{2018PhRvB..98x1412C}, the system is \textit{not} in the diabatic regime of $\omega_{\vect{q}}$\,$\gg$\,$E_F$ \cite{gor2016superconducting,sadovskii2019electron}.

Integrating out the phonon modes, one arrives, within the conventional BCS approximation, at the phonon-mediated electron-electron interaction, which is of the form of a density-density interaction,
\begin{alignat}{1}
\label{eq:Hint}
H^{}_{\mathrm{int}} = \frac{1}{N} \sum_{\vect{q}}  \sum_{l,l'=1}^2\sum_{s,s'=1}^2 V_{l,l'}^{s,s'}(\vect{q})\, \hat{\varrho}^{\pdagger}_{\vect{q},l,s}\, {\hat{\varrho}}^{\pdagger}_{-\vect{q},l',s'},
\end{alignat}
where $\hat{\varrho}^{}_{\vect{q},l,s}$ is the electronic density operator in sheet $s$\,$=$\,$1,2$ of the $l$-th BLG at in-plane momentum $\vect{q}$, i.e., $\hat{\rho}^{}_{\vect{q},l}$\,$=$\,$\sum_{s}\hat{\varrho}^{}_{\vect{q},l,s}$.
For the case of the electron-phonon coupling discussed above, we obtain 
\begin{equation}
V_{l,l'}^{s,s'}(\vect{q}) = -\delta_{l,l'} \frac{D^2}{2\, \bar{m}\, v^2_{ph}\,\Omega} \sum_{n} \frac{\omega^2_{\vect{q}}}{\omega^2_{\vect{q}}+ \nu_n^2}
\end{equation} 
with bosonic Matsubara frequencies $\nu_n$.

In order to study superconducting instabilities, we project this interaction to the partially-filled conduction band harboring superconductivity. Within our approximations, the projected interaction and the noninteracting band structure of the system do not couple the spins of electrons in different valleys.  
This enhances the spin symmetry from the usual $\text{SU}(2)$ to $\text{SU}(2)_+ \times \text{SU}(2)_-$, corresponding to independent spin rotation in the two valleys. In this limit, singlet and triplet are exactly degenerate, which enables us to discuss both of them simultaneously; the type of singlet or triplet pairing that arises when breaking the enhanced spin symmetry is completely determined by symmetry and has been worked out in Ref.~\cite{2019arXiv190603258S}. 
The microscopic features of this projection are captured by the ``form factors'' $F^{l,s}_{\vect{k},\vect{q},\xi}$ involved in the projection of the density operator to the low-energy conduction bands, i.e., 
\begin{equation*}
\hat{\varrho}_{\vect{q},l,s} \rightarrow \sum_{\vect{k}, \sigma, \xi}  \frac{1}{N}c^\dagger_{\vect{k},\sigma,\xi}\,c^\pdagger_{\vect{k}+\vect{q},\sigma,\xi} \, F^{l,s}_{\vect{k},\vect{q},\xi},
\end{equation*}
where $c_{\vect{k},\sigma,\xi}$ annihilates an electron in the conduction band with momentum $\vect{k}$ and spin $\sigma$ in valley $\xi$. In \figref{fig:Phonon}(a), we show the combination of form factors $\sum_{l}|\sum_s F^{l,s}_{\vect{k},\vect{q},+}|^2$ that enters the mean-field equations for phonon-mediated pairing.

The full mean-field theory of superconductivity in TDBG is developed in Appendix~\ref{app:mf}; while we relegate the details to that section, let us highlight here the salient features thereof. Focusing on the energetically most favorable intervalley pairing, the superconducting order parameter $\hat{\Delta}$ becomes a momentum-dependent $2$\,$\times$\,$2$ matrix in spin space that couples as $\propto c^\dagger_{\vect{k},\sigma,+} (\hat{\Delta}_{\vect{k}})^\pdagger_{\sigma\sigma'} c^\dagger_{-\vect{k},\sigma',-}$ to the low-energy electrons. In the  Nambu basis \cite{PhysRev.117.648}, $\psi_{\vect{k},\sigma}^\mathrm{T}$\,$=$\,$(c^{}_{\vect{k},\sigma,+},\, c^{\dagger}_{-\vect{k},\sigma,-} )^\mathrm{T}$, the mean-field Hamiltonian can be expressed in the usual Bogoliubov-de Gennes (BdG) \cite{zhu2016bogoliubov} form as
\begin{equation}\label{BdGFormofHam}
    H^{\textsc{mf}} = \sum_{\vect{k}} \psi^\dagger_{\vect{k},\sigma} 
\begin{bmatrix} 
E^{}_{\vect{k},+} \delta^{}_{\sigma,\sigma'} &  \left(\hat{\Delta}^{\pdagger}_{\vect{k}}\right)^{}_{\sigma, \sigma'}  \\
 \left(\hat{\Delta}^{\dagger}_{\vect{k}}\right)^{}_{\sigma, \sigma'}  & -E^{}_{\vect{k},+} \delta^{}_{\sigma,\sigma'} 
\end{bmatrix}
\psi^\pdagger_{\vect{k},\sigma'}.
\
\end{equation}
Due to the $\text{SU}(2)_+ \times \text{SU}(2)_-$ spin symmetry, singlet, $\hat{\Delta}_{\vect{k}}$\,$=$\,$i\sigma_2 \Delta_{\vect{k}}$, and triplet, $\hat{\Delta}_{\vect{k}}$\,$=$\,$i\sigma_2\, \vect{\sigma}$\,$\cdot$\,$\vect{d}\, \Delta_{\vect{k}}$, are exactly degenerate. Which of the two is realized depends on the sign of the interaction breaking the enhanced spin symmetry, which we come back to later in Sec.~\ref{Interplay} below.
For details on the form of the triplet vector, the admixture of singlet and triplet despite the absence of spin-orbit coupling, and the behavior of the irreducible representation (IR) $E$, we direct the interested reader to \refcite{2019arXiv190603258S}.

For the electron-phonon interaction introduced above, which exhibits an exact $\text{SU}(2)_+ \times \text{SU}(2)_-$ symmetry, we can treat singlet and triplet on equal footing. The momentum dependence of $\Delta_{\vect{k}}$ close to the superconducting transition and, hence, its IR can be inferred from the linearized gap equation,
\begin{subequations}
\begin{alignat}{1}
\label{eq:gapeq1}
\Delta^{}_{\vect{k}} = \sum_{\vect{k}'} \mc{M}^{}_{\vect{k}, \vect{k}'} (T)\, \Delta^{}_{\vect{k}'}, \quad \vect{k} \in \text{FBZ},
\end{alignat}
with a kernel in momentum space given by
\begin{equation}
    \mc{M}^{}_{\vect{k}, \vect{k}'} =\frac{ \mc{V}_{\vect{k},\vect{k}'}}{2 E^{}_{\vect{k},+}} \tanh \left(\frac{E^{}_{\vect{k},+}}{2T} \right), \label{KernelM}
\end{equation}\label{eq:gapeq}\end{subequations}
where $\mc{V}_{\vect{k},\vect{k}'}$ [defined in Eq.~\eqref{eq:v_ref}] is simply a symmetrized rewriting of the interaction $V_{l,l'}^{s,s'}$, modulated with the appropriate form factors.

\subsection{Results}
The linearized mean-field equation \eqref{eq:gapeq}, which has the form of an eigenvalue problem of the kernel $\mathcal{M}(T)\,\Delta$\,$=$\,$\lambda(T)\,\Delta$ in momentum space, is solved on a Monkhorst-Pack grid for the Brillouin zone.
Estimating $D^2/(4 \bar{m}^{}_m v^2_{ph})$\,$\approx$\,$82.3$~meV~nm$^2$ \cite{wu2019phonon, li2019phonon}, we evaluate the critical temperature $T_c$ of superconductivity by determining when the largest of the eigenvalues $\{\lambda_n(T)\}$ of $\mathcal{M}(T)$ reaches $1$; for example, at three-quarters filling, we obtain $T_c$\,$=$\,$1.90\,$K [see Fig.~\ref{fig:Phonon}\,(b)]. Interestingly, this value is of the same order as the experimental observations and suggests that electron-phonon coupling is strong enough to induce superconductivity in TDBG as well. As seen in Fig.~\ref{fig:Phonon}\,(c), $T_c$ is peaked slightly above half filling of the conduction band; this peak in $T_c$ lines up with the maximum of the DOS in Fig.~\ref{fig:band}\,(c), as expected.

More importantly, beyond these numerical values, this formalism enables us to determine the IRs under which the superconducting instabilities transform. The crystalline point group of TDBG is $C_3$. It is well known \cite{dresselhaus2008group} that $C_3$ has two IRs $r$, both
of which are one-dimensional: the trivial one, $A$, and
the complex representation $E$ (and its complex conjugate
partner).
The leading instability is found to transform under the trivial IR, $r$\,$=$\,$A$, while the subleading pairing state transforms under $r$\,$=$\,$E$; the calculated basis functions, $\Delta^r_{\vect{k}\mu}$, are displayed in \figref{fig:Phonon}\,(d). The fact that the dominant basis function respects all symmetries of the system and can be chosen to be real and positive for all momenta in the MBZ is consistent with the general results in \refcite{GeneralProof} (that hold beyond the mean-field approximation). Lastly, we note that all order parameters in \figref{fig:Phonon}\,(d) are neither even nor odd functions of $\vect{k}$, as a consequence of broken inversion and two-fold rotation symmetry in the system.

\section{Superconductivity driven by Coulomb interactions}
\label{sec:ee}

\begin{figure*}[htb]
\includegraphics[width=\linewidth]{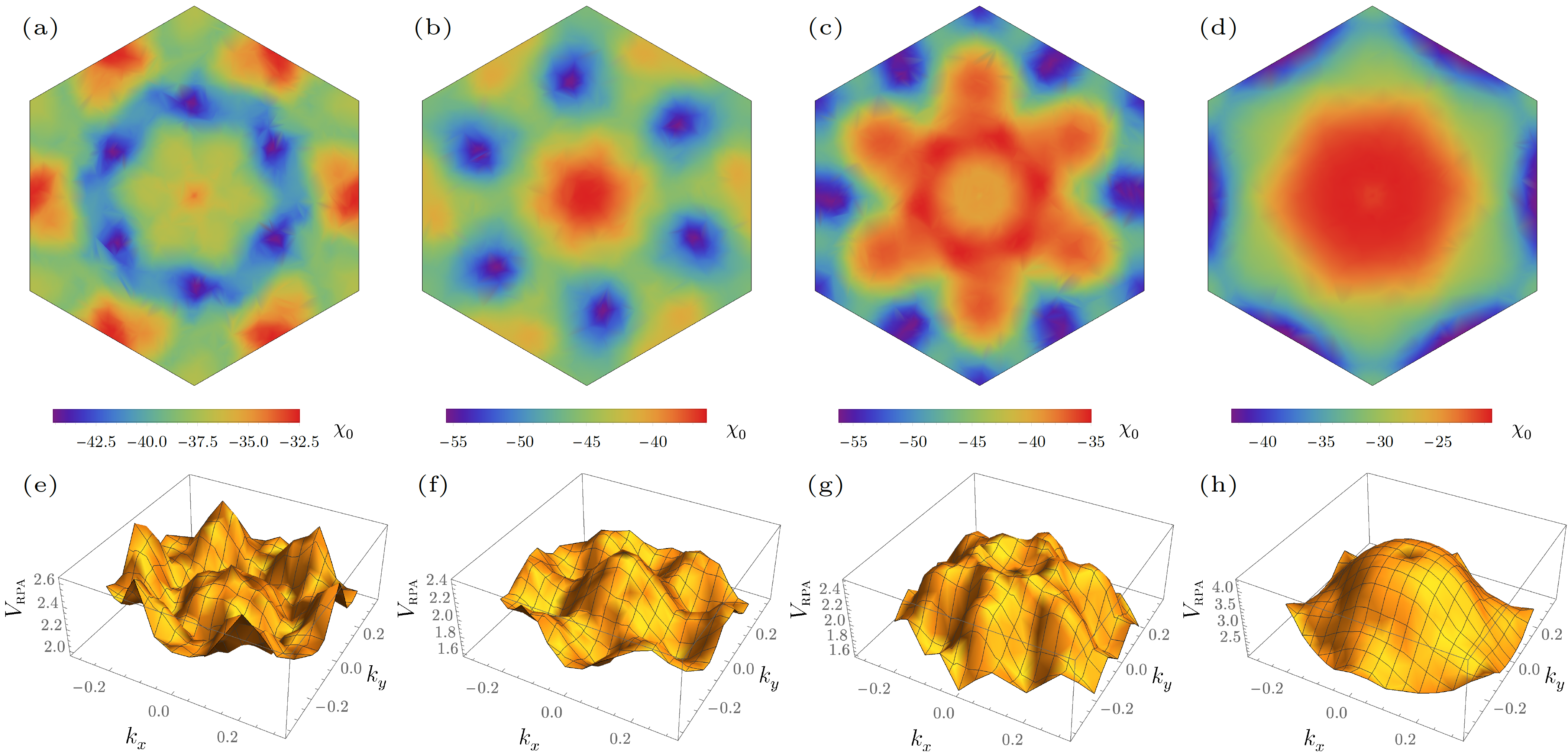}
\caption{\label{fig:RPA}(a--d) The RPA polarization function $\chi^{}_0$ (in keV$^{-1}$\AA$^{-2}$), and (e--h) the renormalized Coulomb potential (in meV) at quarter, one-third, two-thirds, and three-quarters filling [from left to right] of the first conduction band in TDBG at 1.75 K.}
\end{figure*}

On top of reducing the critical temperature in an electron-phonon-driven superconductor, 
the Coulomb repulsion can also be the central driving force of superconductivity \cite{PhysRevLett.15.524, baranov1992superconductivity}, as has, for instance, been discussed in the contexts of monolayer graphene \cite{PhysRevB.78.205431, nandkishore2012chiral,PhysRevB.86.020507,PhysRevB.88.125434}, artificial graphene \cite{2019arXiv190907401L}, and TBG \cite{YiZhuangPairing,PhysRevLett.122.026801}. To explore the scenario of an electronic pairing mechanism in TDBG, we first neglect the electron-phonon coupling and focus on the Coulomb interaction. 
Since the detailed form of the associated interaction $V_{l,l'}^{s,s'}$ in \equref{eq:Hint} at the energy scale of our single-band description---including any potential dependence on the layers $(l,s)$ and $(l',s')$ of the electrons involved---is presently unclear, we assume that it can be described as a pure density-density interaction in the low-energy conduction band: $H_{\text{int}}$\,$=$\,$N^{-1}\sum_{\vect{q}} V(\vect{q}) \rho^c_{\vect{q}}\,\rho^c_{-\vect{q}}$ with $\rho^c_{\vect{q}}$\,$=$\,$\sum_{\vect{k},\sigma,\xi}\,c^\dagger_{\vect{k},\sigma,\xi}\,c^\pdagger_{\vect{k}+\vect{q},\sigma,\xi}$ and \cite{2019arXiv190308685L}
\begin{alignat}{1}
\label{eq:VCoul}
V(\vect{q}) =  \frac{V_0\, \lvert q^{}_\textsc{m} \rvert}{2 \sqrt{\vect{q}^2+\kappa^2}}, \, V_0 = \frac{e^2}{2 \epsilon_0 \epsilon\, \lvert \Omega\, \rvert q^{}_\textsc{m}} \approx \frac{e^2 \theta}{4 \pi \epsilon_0 \epsilon a},
\end{alignat}
where $e$ is the electronic charge and $\epsilon_0$ the vacuum permittivity. 
In \equref{eq:VCoul}, we have already taken into account screening from the gate \cite{papic2014topological, goodwin2019critical}, as described by the screening length $\kappa^{-1}$\,$=$\,$2 \times 10^{-8}~$m \cite{2019arXiv190308685L}, as well as from the surrounding boron nitride and the higher bands of TDBG, captured by the effective dielectric constant $\epsilon$\,$=$\,$5$ \cite{C8NA00350E}. In addition, we next consider the crucial momentum and doping dependence resulting from the internal screening processes of the electrons in the low-energy bands hosting superconductivity.

\subsection{Screening in the random phase approximation}

With the aim of describing screening in TDBG, we now turn to the random phase approximation (RPA) \cite{PhysRev.92.609}. This is primarily motivated by earlier works on TBG, where such an approach has been extensively employed  \cite{ PhysRevB.100.161102,goodwin2019attractive, vanhala2019constrained,2019arXiv190903514W}. In particular, its utility was underscored by Ref.~\cite{YiZhuangPairing}, which discussed the applicability of this method to TBG slightly away from the magic angle---this suggestively bodes well for its extension to TDBG, where there is never a magic angle to begin with, at least in our continuum model. The important caveat to be cognizant of is that RPA does not apply to the strong-coupling regime, and, in principle, more unbiased methods such as the function renormalization group \cite{PhysRevB.99.094521} could be implemented to tackle this problem. Nonetheless, it is encouraging to view that, with regard to electronic ordering instabilities in TBG, RPA  calculations \cite{2018SSCom.282...38L} were found to yield qualitatively similar results to renormalization group methods \cite{2018PhRvX...8d1041I}. Besides, given that we systematically study the pairing states in a weak-coupling single-band description, this simultaneously shows that any deviations from the analysis below (as might eventually be established in future experiments) must result from the strong-coupling or interband nature of superconductivity. 

Using the low-energy continuum model and restricting ourselves to the lowest conduction band as before, we compute the Lindhard function    
\cite{ando2006screening, katsnelson2012graphene}
\begin{alignat}{1}
\Pi^{}_0 (\vect{q}, \omega=0) = \frac{2}{N}\sum_{\vect{k},\xi} \frac{f_{\vect{k},\xi}-f_{\vect{k}+\vect{q},\xi}}{E_{\vect{k},\xi}-E_{\vect{k}+\vect{q},\xi}},
\end{alignat}
where the factor of 2 represents the spin degeneracy, and $f_{\vect{k},\xi}$ is the Fermi function evaluated at band energy $E_{\vect{k},\xi}$. The resultant effective screened electron-electron interaction is given by $V^{}_{\textsc{rpa}}(\vect{q})$\,$=$\, $V (\vect{q})/(1-\Pi^{}_0 (\vect{q}) \,V (\vect{q}))$, which is plotted in Fig.~\ref{fig:RPA}, together with $\chi_0 (\vect{q})$\,$\equiv$\,$\Pi_0 (\vect{q})/\Omega$. The coupling is strongly renormalized 
as a result of this internal screening and exhibits a significant dependence on electronic filling. We can now straightforwardly apply the mean-field machinery established above in Sec.~\ref{ElectronPhononCoupl} to examine the possibility of Coulomb-driven superconductivity. To be precise, Eq.~\eqref{eq:gapeq} still holds; the only change is in the kernel $\mathcal{M}$ in \equref{KernelM} as the phononic interaction $V_{l,l'}^{s,s'}$ has to be replaced with its RPA-screened counterpart $V^{}_{\textsc{rpa}}$.

\subsection{Resultant order parameters}
The screening of the Coulomb interaction is highly anisotropic, leading to a nontrivial texture of $V_{\textsc{rpa}}$ in momentum space. If $V_{\textsc{rpa}}(\vect{q})$\,$>$\,$0\,\forall\,\vect{q}$ and has a local maximum at $\vect{q}$\,$=$\,$0$, one expects that the kernel $\mc{M}(T)$ of the gap equation does not have a positive eigenvalue, signaling the absence of a superconducting instability. This is indeed what we find  below half-filling of the isolated band, i.e., for hole doping relative to the correlated insulator, with $V_{\textsc{rpa}}$ illustrated by Figs.~\ref{fig:RPA}\,(e--f). 
However, as \figref{fig:RPA}\,(g) conveys, the RPA screening leads to a local minimum at $\vect{q}$\,$=$\,$0$ for $2/3$ filling. A superconducting instability now becomes possible as the pairing can gain energy if there is a relative sign change between the order parameters $\Delta_{\vect{k}}$ and $\Delta_{\vect{k}'}$ connected by the momentum $\vect{q}$\,$=$\,$\vect{k}'$\,$-$\,$\vect{k}$\,$\ne$\,$0$ where $V_{\textsc{rpa}}(\vect{q})$ is maximal. In fact, the deepest local minimum is obtained at approximately $\mu$\,$=$\,$0.016$ eV, which roughly coincides with the optimal doping for superconductivity noticed in experiments. Moving away from this filling, on the electron-doped side, the trough quickly disappears [as is visible in Fig.~\ref{fig:RPA}\,(h)], thereby killing any Coulomb-driven superconductivity in the process.

At the electron concentrations where superconductivity is favored, we find attractive interactions both in the $A$ and $E$ representations with basis functions shown in Fig.~\ref{fig:ee}; as in the case of electron-phonon coupling, the $A$ representation is dominant. However, the associated order parameter $\Delta_{\vect{k}}^A$ now exhibits sign changes that are not imposed by the IR of the superconducting state but result from the repulsive nature of the interactions. It displays $6$ of these accidental nodes on the Fermi surface [Fig.~\ref{fig:ee}\,(a)], which is the minimal nonzero number of nodal points for an order parameter transforming as $A$ of $C_3$. An estimate for the superconducting critical temperature with this electronic mechanism yields $T_c$\,$<$\,$0.1\,$K, which is significantly lower than the measured value. Since higher-order corrections to RPA can enhance superconductivity \cite{Efremov2000}, this might well be a consequence of our weak-coupling approach and does not rule out an electronic pairing mechanism.

Despite transforming under the same, trivial, representation, the dominant superconducting states in \figref{fig:Phonon}\,(d) and Fig.~\ref{fig:ee}\,(a) for phonon-mediated and electronic pairing mechanisms, respectively, will have very different thermodynamic properties due to the absence and presence of accidental nodes.

\subsection{Interplay of pairing mechanisms and breaking of $\text{SU}(2)_+ \times \text{SU}(2)_-$ symmetry}\label{Interplay}
So far, we have focused on interactions with an exact $\text{SU}(2)_+ \times \text{SU}(2)_-$ spin symmetry, rendering singlet and triplet degenerate. We will here consider different ways of breaking this symmetry and discuss which type of pairing will be favored. We emphasize that the following results are not limited to TDBG, but are more generally relevant to other graphene moir\'e superlattice systems as well.

We first point out that singlet will generically dominate over triplet if phonon-mediated interactions alone are responsible for pairing \cite{GeneralProof,PhysRevB.90.184512}. 
If, however, the $\text{SU}(2)_+ \times \text{SU}(2)_-$-symmetry-breaking interactions are dominated by the Coulomb repulsion, triplet pairing is possible; this is true even when the main pairing glue is the $\text{SU}(2)_+ \times \text{SU}(2)_-$ symmetric electron-phonon-induced interaction---a plausible assumption for graphene moir\'e systems given our analysis of electron-phonon pairing presented above, and the additional insights from experiments on TBG that reveal the relevance of phonons for pairing \cite{2019arXiv191109198S,2019arXiv191113302S}.

To study this scenario further, let us assume that the dominant contribution to the $\text{SU}(2)_+ \times \text{SU}(2)_-$-symmetry breaking interactions comes from the fluctuation of a set of collective electronic modes, $\phi_{\vect{q}}^j$, that couple to the electrons, $c_{\vect{k},\alpha}$, according to
\begin{equation}
    H^\pdagger_{c\phi} = \sum_{\vect{k},\vect{q}} c_{\vect{k}+\vect{q},\alpha}^\dagger \lambda^{j}_{\alpha\beta}(\vect{k}+\vect{q},\vect{k})c_{\vect{k},\beta}^\pdagger \, \phi_{\vect{q}}^{j}. \label{GeneralFermionBosonCoupling}
\end{equation}
Here, we allow for a momentum-dependent coupling matrix $\lambda$ that can couple different internal degrees of freedom, captured by the multi-indices $\alpha$, $\beta$ (spin, valley, and potentially also different bands). We show in Appendix~\ref{app:singlettriplet} that, irrespective of the detailed form of $\lambda$ and the action of the collective mode $\phi_{\vect{q}}^{j}$, the competition between singlet and triplet is determined by the properties of $\phi_{\vect{q}}^{j}$ under time reversal: if $\phi_{\vect{q}}^{j}$ is time-reversal even (odd), i.e., it corresponds to a nonmagnetic (magnetic) particle-hole order parameter such as charge (spin) fluctuations, singlet (triplet) will dominate. Due to the proximity of the superconductor in graphene moir\'e systems to magnetic phases \cite{2019arXiv190306952S,ExperimentKim,PabllosExperiment,Sharpe605,2019arXiv190306513L,FMTrilayer}, this reveals that these systems can harbor a triplet pairing phase, even when the pairing glue is provided by the electron-phonon coupling.

\section{Disorder sensitivity}
\label{sec:disorder}

\begin{figure}[t]
\includegraphics[width=\linewidth]{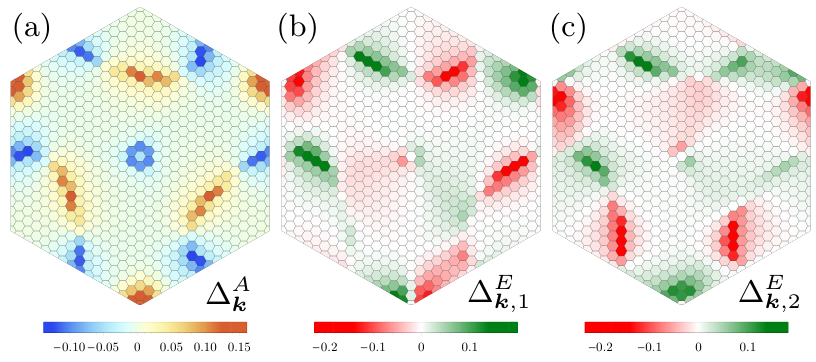}
\caption{\label{fig:ee}Profiles of the different order parameters that transform under the $A$ (dominant channel) and $E$ (subleading) representations when superconductivity is mediated by the repulsive Coulomb interaction alone.}
\end{figure}

In this section, we analyze the disorder sensitivity of the dominant superconducting states derived in Secs.~\ref{sec:eph} and \ref{sec:ee} above for electron-phonon and Coulomb-driven pairing. We will see that although both transform under the IR $A$, they behave quite differently in the presence of impurities. In addition, we discuss an ``Anderson theorem'' for triplet pairing that naturally emerges in graphene-based moir\'e superlattice systems as a result of the valley degree of freedom.

In Ref.~\cite{2020arXiv200104673T}, a general expression for the disorder-induced suppression, $\delta T_c$\,$=$\,$T_c - T_{c,0}$, of the transition temperature $T_c$ of a superconductor with respect to its clean value $T_{c,0}$ was derived: for weak scattering and/or low disorder concentrations, $\tau^{-1} \rightarrow 0$, it holds that
\begin{equation}
\label{eq:reduction}
    \delta T_{c}\sim -\frac{\pi}{4}\tau^{-1} \, \zeta.
\end{equation}
All nonuniversal details of the system and the impurity potential are encoded in the sensitivity parameter $\zeta$ that can be written as a Fermi-surface average of the trace $\text{tr}[C^\dagger_{\vect{k},\vect{k}'}C^\pdagger_{\vect{k},\vect{k}'}]$ where $C^{}_{\vect{k},\vect{k'}}$ is the (anti)commutator
\begin{equation}
    C^{}_{\vect{k},\vect{k'}} =  \widetilde{\Delta}^{}_{\vect{k}} \mathscr{T}^\dagger W^{}_{\vect{k},\vect{k}'} - t^{}_W W^{}_{\vect{k},\vect{k}'}\widetilde{\Delta}^{}_{\vect{k}'}\mathscr{T}^\dagger.  \label{FormOfTheCommutator}
\end{equation}
Here, $\mathscr{T}$ is the unitary part of the time-reversal operator (in our case $i\sigma_2 \tau_1$ with Pauli matrices $\sigma_j$ and $\tau_j$ in spin and valley space, respectively), $W_{\vect{k},\vect{k}'}$ is the Fourier transform of the impurity potential (in our low-energy description, a matrix in spin and valley space), and $t_W = +1$ ($t_W = -1$) for nonmagnetic (magnetic) disorder. Furthermore, $\widetilde{\Delta}_{\vect{k}}$ is the superconducting order parameter, now expressed in the full valley-spin space, i.e., we have $\widetilde{\Delta}_{\vect{k}}^\pdagger\mathscr{T}^\dagger = \Delta_{\vect{k}}\, \sigma^\pdagger_0 \tau^\pdagger_0$ and $\widetilde{\Delta}^\pdagger_{\vect{k}}\mathscr{T}^\dagger = \Delta_{\vect{k}}\, \vect{\sigma}\cdot \vect{d}\, \tau^\pdagger_3$ for singlet and triplet pairing, respectively.
As follows from our analysis above, the leading superconducting instabilities have $\Delta_{\vect{k}}= \Delta_{\vect{k}}^A$, transforming trivially under the point group $C_3$ of the system (IR $A$), and its explicit form is given by the left panel in Fig.~\ref{fig:Phonon}\,(d), for electron-phonon pairing, and by Fig.~\ref{fig:ee}\,(a) for Coulomb-driven superconductivity.

To be concrete, let us focus on spinless, nonmagnetic ($t_W$\,$=$\,$+1$) local impurities, i.e., $W_{\vect{k},\vect{k}'} = \sigma_0(w_1 \tau_0 + w_2 \tau_{1} + w_3 \tau_2)$, $w_j \in \mathbb{R}$, normalized such that $\sum_j w_j^2 = 1$. As in Ref.~\cite{2020arXiv200104673T}, the normalization is chosen so as to ensure that spin-magnetic disorder, $W_{\vect{k},\vect{k}'} = \sigma_3(w_1 \tau_0 + w_2 \tau_{1} + w_3 \tau_2)$, yields $\zeta=1$ for a momentum-independent singlet order parameter, i.e., in the Abrikosov-Gorkov limit \cite{AGLaw}.

The sensitivity parameter can be written as \cite{2020arXiv200104673T}
\begin{equation}
    \zeta =  \frac{\sum_{\vect{k},\vect{k}'}^{\text{FS}}\text{tr}\left[ C^\dagger_{\vect{k},\vect{k}'}C^\pdagger_{\vect{k},\vect{k}'} \right] }{4\sum_{\vect{k}}^{\text{FS}}\text{tr}\left[\widetilde{\Delta}_{\vect{k}}^\dagger \widetilde{\Delta}_{\vect{k}}^\pdagger\right]}, \label{ExpressionForZeta}
\end{equation}
where $\sum_{\vect{k}}^{\text{FS}} \mc{F}_{\vect{k}} \equiv \braket{\mc{F}_{\vect{k}}}_{\text{FS}}$ represents an average of a function $\mc{F}_{\vect{k}}$ over the Fermi surfaces of the system (normalized such that $\braket{1} = 1$). 
After some algebra, we are led to
\begin{equation}
    \zeta = \frac{\braket{|\Delta_{\vect{k}}^A|^2}_{\text{FS}} - |\braket{\Delta_{\vect{k}}^A}_{\text{FS}}|^2}{2 \braket{|\Delta_{\vect{k}}^A|^2}_{\text{FS}}} \label{SingletStateZeta}
\end{equation}
for singlet pairing. We see that, independent of the ratio of intra- ($w_1$) to intervalley ($w_2$ and $w_3$) scattering, the sensitivity of the singlet state is just the variance of the basis function $\Delta_{\vect{k}}^A$ on the Fermi surface. For the basis function of the conventional pairing state illustrated in the first panel of \figref{fig:Phonon}\,(d), the variance is small, so one would expect $\zeta \ll 1$. Evaluating \equref{SingletStateZeta} with the explicit form of our order parameter, we indeed find a small value, $\zeta \approx 0.058$, and the state is quite robust against nonmagnetic disorder, similar to the conventional ``Anderson theorem'' \cite{AT,ATAG1}. 

Contrarily, despite transforming under the trivial representation, the order parameter in  \figref{fig:ee}\,(a) has many sign changes and hence, a small value of $\braket{\Delta_{\vect{k}}^A}_{\text{FS}}$ compared to $\braket{|\Delta_{\vect{k}}^A|}_{\text{FS}}$. Accordingly, $\zeta$ should be close to the maximal possible value of $1/2$ that is typically associated only with nontrivial representations, where $\braket{\Delta_{\vect{k}}^r}_{\text{FS}} = 0$ by symmetry. From the momentum dependence in Fig.~\ref{fig:ee}\,(a) we obtain $\zeta \approx 0.498$ and the state is fragile against disorder. This---or the subtle doping dependence of the electronic pairing mechanism in \figref{fig:RPA}---among other reasons, could make superconductivity challenging to observe \cite{PabllosExperiment,burg2019correlated}.

Next, we analyze triplet pairing. From \equref{ExpressionForZeta}, we find, 
\begin{equation}
    \zeta = \frac{\braket{|\Delta_{\vect{k}}^A|^2}_{\text{FS}} - (w_1^2-w_2^2-w_3^2) |\braket{\Delta_{\vect{k}}^A}_{\text{FS}}|^2}{2 \braket{|\Delta_{\vect{k}}^A|^2}_{\text{FS}}}.
\end{equation}
Consequently, if intravalley scattering dominates, i.e., $w_1$\,$\gg$\,$ w_{2,3}$, the triplet and singlet states behave identically. Intuitively, this results from the fact that the relative sign change of the triplet order parameter between the two valleys is irrelevant if intervalley scattering is weak. Note that the values of $w_{1,2,3}$ parametrizing the impurity potential are immaterial for the basis function in Fig.~\ref{fig:ee}\,(a) for which, $|\braket{\Delta_{\vect{k}}^A}_{\text{FS}}|$\,$\ll$\,$\braket{|\Delta_{\vect{k}}^A|}_{\text{FS}}$, so $\zeta$\,$\approx$\,$0.5$, as in the singlet case. The only difference between singlet and triplet arises for the phonon-mediated state, where $\braket{|\Delta_{\vect{k}}^A|^2}_{\text{FS}}$\,$\approx$\,$|\braket{\Delta_{\vect{k}}^A}_{\text{FS}}|^2$, leading to $\zeta$\,$\approx$\,$w_2^2 + w_3^2$. Here, the stability of superconductivity is determined by the ratio of intra- to intervalley pairing. 

While, in general, impurities can mediate scattering between different valleys, the intervalley scattering amplitudes will be suppressed as long as the impurity potential only varies on the length scales of the moir\'e lattice, but is smooth on the atomic scale. Thus, the triplet state with basis function in Fig.~\ref{fig:Phonon}\,(d), which can be stabilized by a combination of electron-phonon and electron-electron coupling (see Sec.~\ref{Interplay}), can also exhibit an analogue of the Anderson theorem and be protected against nonmagnetic disorder on the moir\'e scale. To see this explicitly, let us start from the tight-binding description of the individual graphene layers. We denote the annihilation operator of an electron on Bravais lattice site $\vect{R}$, sublattice $\tau$, spin $\sigma$, and layer $\ell$ by $a^\dagger_{\vect{R},\sigma, \ell, \tau}$ and consider spinless impurities of the general form
\begin{equation}
    H_{\text{imp}} = \sum_{\vect{R}}a^\dagger_{\vect{R},\sigma, \ell, \tau}\, v_{\tau\tau'}^{\ell \ell'}(\vect{R})\, a^\pdagger_{\vect{R},\sigma, \ell', \tau'}.
\end{equation}
In order to connect to the continuum-model operators, e.g., in \equref{eq:density}, we first Fourier transform to get
\begin{equation}
    H_{\text{imp}} = \sum_{\vect{k},\vect{q}}a^\dagger_{\vect{k},\sigma, \ell, \tau}\, \widetilde{v}_{\tau\tau'}^{\ell \ell'}(\vect{q}) \,a^\pdagger_{\vect{k}+\vect{q},\sigma, \ell', \tau'}, \label{FourierTransformOfImpurityPotential}
\end{equation}
where $\widetilde{v}_{\tau\tau'}^{\ell \ell'}(\vect{q}) = N^{-1} \sum_{\vect{R}} v_{\tau\tau'}^{\ell \ell'}(\vect{R}) e^{i \vect{R}\cdot \vect{q}}$ is the Fourier transform of the impurity potential. Clearly, if the potential of the local perturbation is smooth on the atomic length scales, $\widetilde{v}_{\tau\tau'}^{\ell \ell'}(\vect{q})$ with $\vect{q}$ connecting the $K$ and $K'$ points (in the original Brillouin zone of graphene) is negligibly small. Transforming back to real space, \equref{FourierTransformOfImpurityPotential} can, thus, be approximately written as
\begin{equation}
    H_{\text{imp}}  \approx \int \diff \vect{r}\, a^\dagger_{\sigma, \ell, \tau, \xi} (\vect{r})\,v_{\tau\tau'}^{\ell \ell'}(\xi;\vect{r}) \,a^\pdagger_{\sigma, \ell', \tau', \xi} (\vect{r}),
\end{equation}
in the continuum description.
Most importantly, it is diagonal in valley space. This property will remain unchanged upon projection onto the bands of the system (see \appref{app:ff}). 

We finally note that the resulting protection of the triplet state against this type of disorder is not altered by the fact that, in general, the impurity potential will be momentum dependent, $W_{\vect{k},\vect{k}'} = \tau_0\sigma_0 f^{(1)}_{\vect{k},\vect{k}'}+\tau_3\sigma_0 f^{(2)}_{\vect{k},\vect{k}'}$ with scalar functions $f^{(1,2)}$ (constrained to respect time-reversal symmetry): 
as long as the basis function $\Delta_{\vect{k}}^A$ is (almost) momentum independent, the commutator in \equref{FormOfTheCommutator} will vanish (approximately), wherefore $\zeta=0$ ($\zeta \ll 1$).\\


\section{Conclusion and outlook}
\label{sec:end}
In this work, we have analyzed the dominant and subleading superconducting instabilities for TDBG using a microscopic continuum model, and discussed both electron-phonon and RPA-screened purely electronic pairing mechanisms. In both cases, the leading superconducting instability transforms under the IR $A$, i.e., is invariant under all point symmetries of the system, and its order parameter can be found in \figref{fig:Phonon}(d), left panel, and \figref{fig:ee}(a), respectively. We expect the explicit form of the derived superconducting order parameters to be useful for comparison with future experimental investigations of the system such as quasiparticle-interference experiments. On the theoretical side, further first-principles computations of phononic properties \cite{cocemasov2013phonons, li2014thermal} of TDBG are needed to obtain a refined description of electron-phonon pairing.

We also discussed the scenario that the dominant pairing glue, which preserves the enhanced $\text{SU}(2)_+ \times \text{SU}(2)_-$ spin symmetry of the system, is provided by the electron-phonon coupling while the residual interactions breaking this symmetry are associated with fluctuating bosonic modes $\phi^j$. It is shown that time-reversal even (odd) $\phi^j$ generically favor singlet (triplet) pairing. In combination with the fact that there are indications of magnetism in the phase diagram of several graphene-based moir\'e superlattice systems \cite{2019arXiv190306952S,ExperimentKim,PabllosExperiment,Sharpe605,2019arXiv190306513L,FMTrilayer}, triplet pairing seems like a natural possibility, even if electron-phonon coupling provides the key ingredient for the value of the critical temperature. In such a scenario, it is possible that the superconducting transition temperature of the triplet is only weakly affected by a reduction of the Coulomb interaction, which could, however, induce a transition from triplet to singlet pairing. We have also demonstrated that, contrary to common wisdom, the associated triplet state will enjoy protection from an Anderson theorem against nonmagnetic impurities at the moir\'e length scales.

\begin{acknowledgments}
We thank Darshan Joshi, Peter Orth, Subir Sachdev, Harley Scammell, and Yanting Teng for valuable discussions. This research was supported by the National Science Foundation under Grant No.~DMR-1664842, and the computing resources for the numerics in this paper were kindly provided by Subir Sachdev. 
\end{acknowledgments}

\begin{appendix}

\section{Continuum model for TDBG}
\label{app:model}
In this section, we briefly review the continuum model for TDBG introduced by \citet{koshino2019band}. Each unrotated bilayer graphene (BLG) sheet has lattice vectors $\vect{a}_1$\,$=$\,$a(1,0)$ and $\vect{a}_2$\,$=$\,$a(1/2,\sqrt{3}/2)$, with lattice constant $a \approx 0.246\,\mathrm{nm}$; their reciprocal lattice counterparts 
are $\vect{b}_1 = (2\pi/a)(1,-1/\sqrt{3})$ and $\vect{b}_2=(2\pi/a)(0,2/\sqrt{3})$.
Once a relative rotation is applied between the two BLGs, the lattice vectors differ from the untwisted case, and, for the $l$-th BLG, are given by $\vect{a}_i^{(l)}$\,$=$\,$R(\mp \theta/2)\vect{a}_i$ 
(with $\mp$ for $l=1,2$), $R (\theta)$ being the matrix for rotations by angle $\theta$. Correspondingly, the reciprocal lattice vectors are modified to $\vect{b}_i^{(l)}$\,$=$\,$R(\mp \theta/2)\,\vect{b}_i$. The moir\'{e} pattern thus formed is characterized in the limit of small $\theta$ by the reciprocal lattice vectors
$ \vect{G}^{\rm M}_{i}$\,$=$\,$\vect{b}^{(1)}_i$\,$-$\,$\vect{b}^{(2)}_i \, (i=1,2)$.
In valley $\xi=\pm 1$, the Dirac points of graphene are located at $\vect{K}^{(l)}_\xi = -\xi\, [2\vect{b}^{(l)}_1+\vect{b}^{(l)}_2]/3$ 
for the $l$-th BLG.

Denoting by $A_{\ell},B_\ell$ the sublattice on layer $\ell=1,2,3,4$ [labeled by the double-index $(l,s)$ in Sec.~\ref{sec:eph}], the continuum Hamiltonian for TDBG at small twist angles $\theta\, (\ll 1)$ can be expressed in the Bloch basis of carbon's $p_z$ orbitals, $(A_1,B_1,A_2, B_2, A_3, B_3, A_4, B_4)$ as
 \begin{align}
&	
{H}^{}_{\textrm{AB-AB}} = 
	\begin{pmatrix}
		H^{}_0(\vect{k}_1) & S^\dagger(\vect{k}_1) & & \\
		S(\vect{k}_1) & H'_0(\vect{k}_1) & U^\dagger  &\\
		& U & H^{}_0(\vect{k}_2) & S^\dagger(\vect{k}_2)  \\
		& & S(\vect{k}_2) & H'_0(\vect{k}_2) \\
	\end{pmatrix}  + V,
	\label{eq_AB-AB}
\end{align}
where $ \vect{k}_l = R(\pm \theta/2)({\vect{k}}-\vect{K}^{(l)}_\xi)$
with $\pm$ for $l=1,2$. Using the shorthand $k_\pm = \xi k_x \pm i k_y$, the other building blocks are
\begin{align}
\nonumber  H^{}_0(\vect{k}) 
&=
\begin{pmatrix}
0  & -\hbar v k_- \\
-\hbar v k_+ & d'
\end{pmatrix},
\,
H'_0(\vect{k}) 
=
\begin{pmatrix}
d'  & -\hbar v k_- \\
-\hbar v k_+ & 0
\end{pmatrix}, 
\\ 
S(\vect{k}) 
&=
\begin{pmatrix}
\hbar v_4 k_+  & \gamma_1 \\
\hbar v_3 k_-  & \hbar v_4 k_+
\end{pmatrix}.
\end{align}
$H_0$ and $H'_0$ above are the Hamiltonians of monolayer graphene, where the band velocity $v$ is $\hbar v /a = 2.1354\,$eV \cite{moon2013optical,koshino2018maximally}.
At so-called \textit{dimer} sites, the $A_1$ atoms of the first layer sit atop the $B_2$ atoms of the second, and this results in the small additional on-site potential $d' = 0.050\,$eV \cite{mccann2013electronic} with respect to nondimer sites.
The interlayer coupling of AB-stacked BLG is captured by the matrix $S$, wherein $\gamma_1 = 0.4\,$eV is the coupling between the abovementioned dimer sites, while the parameters $v_3$ and $v_4$ are related by $v_i = (\sqrt{3}/2) \gamma_i a /\hbar \, (i=3,4)$ \cite{mccann2013electronic} to the diagonal hopping elements $\gamma_3 = 0.32\,$eV and $\gamma_4=0.044\,$eV. In AB-stacked BLG, $v_3$ and $v_4$ account for trigonal warping of the energy band and electron-hole asymmetry, respectively.

The long-wavelength moir\'{e} potential stemming from the angular twist between the two BLGs also induces couplings between the atoms in the second and third layers. This moir\'{e} interlayer hopping is effectively described by the matrix $U$ in Eq.~\eqref{eq_AB-AB}, which is given by
\cite{bistritzer2011moire,moon2013optical,koshino2018maximally}
\begin{align}
 U &= 
\begin{pmatrix}
u & u'
\\
u' & u
\end{pmatrix}
+
\begin{pmatrix}
u & u'\omega^{-\xi}
\\
u'\omega^\xi & u
\end{pmatrix}
e^{i\, \xi \,\vect{G}^{\rm M}_1\cdot\vect{r}}
\nonumber\\
& 
 +
\begin{pmatrix}
u & u'\omega^\xi
\\
u'\omega^{-\xi} & u
\end{pmatrix}
e^{i \,\xi\, (\vect{G}^{\rm M}_1+\vect{G}^{\rm M}_2)\cdot\vect{r}}; \quad\omega =e^{2\pi i/3},
\label{eq_interlayer_matrix}
\end{align}
where $u = 0.0797$ eV and $u' = 0.0975$ eV \cite{koshino2018maximally}; crucially, $u \ne u'$.
This difference between the diagonal ($u$) and off-diagonal ($u'$) amplitudes represents the out-of-plane corrugation effect \cite{koshino2019band}: such lattice relaxation not only expands (shrinks) AB (AA) stacking regions but also enhances the energy gaps separating the lowest-energy and excited bands \cite{koshino2018maximally,nam2017lattice,tarnopolsky2019origin}. In momentum space, the coupling $U$ hybridizes the eigenstates at a Bloch vector $\vect{k}$ in the moir\'{e} Brillouin zone with those at $\vect{q} = \vect{k} + \vect{G}$, where $\vect{G} = m_1  \vect{G}^{\rm M}_1 + m_2  \vect{G}^{\rm M}_2$ for $m_1, m_2 \in \mathbb{Z}$.
 
 \begin{figure}[tb]
\includegraphics[width=\linewidth]{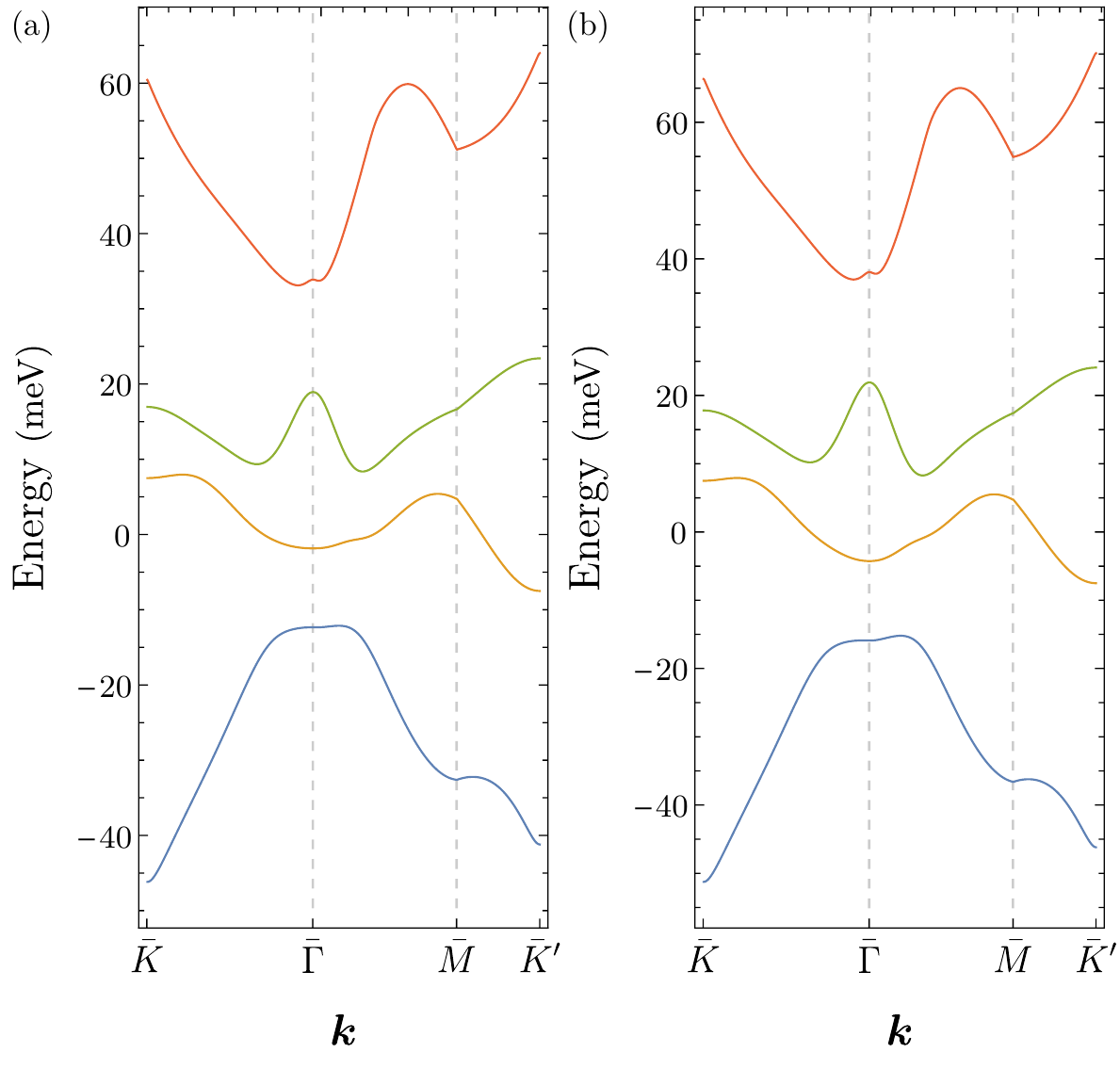}
\caption{\label{fig:band_extra}Band structure of TDBG at (a) $\theta=1.28^\circ$, and (b) $\theta=1.33^\circ$, which are the twist angles used in the devices where \citet{2019arXiv190306952S} observed signatures of superconducting behavior. Note the overall similarity of the dispersions to that in Fig.~\ref{fig:band}(a) at $\theta=1.24^\circ$, for which superconductivity was reported by \citet{ExperimentKim}.}
\end{figure}

Lastly, another important tuning knob in the experimental setup is the potential difference between the top and bottom graphene layers. This lends another useful degree of controllability to the system, enabling one to change the band separations by adjusting the gate voltage difference.
In Eq.~\eqref{eq_AB-AB}, it is parametrized by $V$, the interlayer asymmetric potential,
\begin{alignat}{1}
V = \mathrm{diag} \left(\frac{3\,d}{2}\, \mathds{1}, 
\,\, \frac{d}{2}\,\mathds{1}, 
\,\,-\frac{d}{2}\, \mathds{1}, 
\,\, -\frac{3\,d}{2}\, \mathds{1} \right),
\end{alignat}
where $\mathds{1}$ stands for the $2\times 2$ unit matrix, and $d$ connotes the electrostatic energy difference between adjacent layers, arising from a constant perpendicular field.

Putting all these ingredients together, we numerically diagonalize the Hamiltonian~\eqref{eq_AB-AB} in momentum space (with only a limited number of wavevectors $\vect{q}$ inside the cutoff circle $\lvert \vect{q}- (\vect{K}^{(1)}_\xi+ \vect{K}^{(2)}_\xi)/2\rvert < 4 |\vect{G}^{\rm M}_1|$) to obtain the low-energy eigenstates and energy bands. Despite this truncation, one can ensure satisfactory convergence of the electronic property of interest by choosing a sufficiently large cutoff parameter \cite{PhysRevResearch.1.013001}.  
Since the intervalley coupling is negligible at small twist angles, this calculation can be carried out separately for each valley. Samples of the band structures obtained in this fashion are presented in Figs.~\ref{fig:band}(a) and \ref{fig:band_extra}.

\section{Derivation of the form factors}
\label{app:ff}

In TDBG, there are four main degrees of freedom for our continuum fields, which are indexed by $\sigma$ (spin), $\ell = (l,s)$ (layer), $\tau$ (sublattice), and $\xi$ (valley), so the electronic creation operators at (continuum) position $\vect{r}\in\mathbb{R}^2$ will be denoted by $a_{\sigma, \ell, \tau, \xi}(\vect{r})$. Although the following framework can be applied to any observable, let us begin with the total electronic density at a two-dimensional position $\vect{r}$, given by
\begin{alignat}{1}
\label{eq:density}
\hat{\rho}\,(\vect{r}) = \sum_{\sigma, \ell, \tau, \xi} a^\dagger_{\sigma, \ell, \tau, \xi} (\vect{r}) \,a^\pdagger_{\sigma, \ell, \tau, \xi} (\vect{r})  \equiv \sum_{\alpha} a^\dagger_\alpha(\vect{r})  \,a^\pdagger_\alpha(\vect{r}),
\end{alignat}
where $\alpha$ is a multi-index encompassing all four individual indices. Using the completeness relation
\begin{alignat}{1}
\rvert \alpha, \vect{r} \rangle = \sum_{n} \sum_{\vect{k}}^{\text{FBZ}}  \left \lvert \Psi_{\vect{k},n} \right \rangle \left \langle \Psi_{\vect{k},n}\, \vert\, \alpha, \vect{r} \right \rangle
\end{alignat}
for the Bloch states $\ket{\Psi_{\vect{k},n}}$---where the sum involves all momenta of the first moir\'{e} Brillouin zone (FBZ) and all bands $n$ of the system---we can write, at the level of the field operators
\begin{alignat}{1}
\label{eq:c}
a^\pdagger_\alpha (\vect{r}) = \frac{1}{\sqrt{N}} \sum_{n} \sum_{\vect{k}}^{\text{FBZ}} \left(\vec{\Psi}^{}_{\vect{k},n} (\vect{r}) \right)_\alpha c^\pdagger_{\vect{k},n}.
\end{alignat}
In this equation, we have introduced the vector notation $(\vec{\Psi}^{}_{\vect{k},n} (\vect{r}))_\alpha$\,$\equiv$\,$\braket{\alpha,\vect{r}|\Psi_{\vect{k},n}}$, and $c_{\vect{k},n}$ are the electronic operators in band $n$ with momentum $\vect{k}$. 

When all bands are taken into account, Eq.~\eqref{eq:c} is just a change of basis and, hence, exact. However, in the following we will project the theory on the single conduction band (per spin and valley)---colored green in Fig.~\ref{fig:band}(a)---that hosts superconductivity; within the current notation, this subspace is associated with four values of the band index $n$ that we will, from now on, relabel as $n \rightarrow (\sigma,\xi)$. The projection amounts to restricting the sum on the right-hand side of \equref{eq:c} to only these four values.
Plugging this truncated form of Eq.~\eqref{eq:c} into Eq.~\eqref{eq:density}, the projected electronic density operator, $\rho\,(\vect{r})$, reads as  
\begin{alignat*}{1}
\rho\,(\vect{r}) &=  \frac{1}{N}\sum_{\vect{k}, \vect{k}'} \sum_{\sigma,\sigma',\xi,\xi'} \vec{\Psi}^\dagger_{\vect{k},\sigma,\xi} (\vect{r})\, \vec{\Psi}^\pdagger_{\vect{k}',\sigma',\xi'} (\vect{r})\, c^\dagger_{\vect{k},\sigma,\xi}\,c^\pdagger_{\vect{k}',\sigma',\xi'}.
\end{alignat*}
Recognizing that $\vec{\Psi}^\dagger_{\vect{k},\sigma,\xi} (\vect{r})\, \vec{\Psi}^\pdagger_{\vect{k}',\sigma',\xi'} (\vect{r})$\,$\propto$\,$\delta^{}_{\sigma,\sigma'}$ and independent of $\sigma$, the total density can be split into two components as
\begin{alignat}{1}
\nonumber {\rho}\, (\vect{r}) &= {\rho}^{\coI} (\vect{r}) +  {\rho}^{\coII} (\vect{r}), \mbox{ with }\\
{\rho}^{\coI}\, (\vect{r}) &= \frac{1}{N}\sum_{\vect{k}, \vect{k}',\sigma,\xi} \vec{\Psi}^\dagger_{\vect{k},\xi} (\vect{r})\, \vec{\Psi}^\pdagger_{\vect{k}',\xi} (\vect{r})\, c^\dagger_{\vect{k},\sigma,\xi}\,c^\pdagger_{\vect{k}',\sigma,\xi},\\
\nonumber {\rho}^{\coII}\, (\vect{r}) &= \frac{1}{N}\sum_{\vect{k}, \vect{k}',\sigma,\xi} \vec{\Psi}^\dagger_{\vect{k},\xi} (\vect{r})\, \vec{\Psi}^\pdagger_{\vect{k}',-\xi} (\vect{r})\, c^\dagger_{\vect{k},\sigma,\xi}\,c^\pdagger_{\vect{k}',\sigma,-\xi}.
\end{alignat}
Within the continuum description used in this work, the overlap between the Bloch states for different $\xi$ and, hence, ${\rho}^{\coII}$, vanishes identically. This eventually gives rise to the enhanced $\text{SU}(2)_+ \times \text{SU}(2)_-$ spin-rotation symmetry. In reality, ${\rho}^{\coII}$ is nonzero (but small), which breaks this symmetry. However, as has been shown in \refcite{2019arXiv190603258S}, the behavior when this symmetry is weakly broken is completely determined by symmetry, which is why we do not have to consider it separately in this work.

Once again turning to Bloch's theorem, we can write
\begin{equation}
\vec{\Psi}^{}_{\vect{k},\xi} (\vect{r}) = e^{i \vect{k}.\vect{r}}\, \vec{\mathbb{U}}^{}_{\vect{k},\xi} (\vect{r}) = e^{i \vect{k}.\vect{r}} \sum_{\vect{G}} \vec{\mc{U}}^{}_{\vect{k},\xi} (\vect{G})\, e^{i \vect{G}.\vect{r}}, 
\end{equation}
where $\vec{\mathbb{U}}_{\vect{k},\xi}$ is lattice periodic, but with the periodicity of the moir\'{e} superlattice. The Fourier transform of the density is simply
\begin{alignat}{1}
&{\rho}^\pdagger_{\vect{q}} = \int \mathrm{d} \vect{r}\, e^{-i \vect{q}.\vect{r}} {\rho}\, (\vect{r}) \\
\nonumber &= \frac{1}{N}\sum_{\vect{k}, \vect{k}'}^{\mathrm{FBZ}}  \int \mathrm{d} \vect{r}\hspace*{-0.25cm}
 \sum_{\vect{G}, \vect{G}',\sigma,\xi} \Bigg[\,\vec{\mc{U}}^\dagger_{\vect{k}, \xi} (\vect{G})\,\, \vec{\mc{U}}^\pdagger_{\vect{k}',\xi} (\vect{G'}) \,c^\dagger_{\vect{k},\sigma,\xi}\,c^\pdagger_{\vect{k}',\sigma,\xi}\\
\nonumber &\qquad\qquad \qquad \qquad \quad\,\, \times \exp \left(i (\vect{k}'-\vect{k}+\vect{G}'-\vect{G}-\vect{q}).\vect{r}\right) \Bigg]\\
\nonumber&= \frac{1}{N} \sum_{\vect{k},\vect{G}',\sigma, \xi} c^\dagger_{\vect{k},\sigma,\xi}\,c^\pdagger_{\vect{k}+\vect{q}-\vect{G}',\sigma,\xi} \left[\sum_{\vect{G}}\, \vec{\mc{U}}^\dagger_{\vect{k},\xi} (\vect{G})\,\, \vec{\mc{U}}^\pdagger_{\vect{k} + \vect{q},\xi} (\vect{G}) \right],
\end{alignat}
such that $\vect{k}$, $\vect{k}+\vect{q}-\vect{G}'\in\mathrm{FBZ}$. This condition is satisfied by exactly one $\vect{G}'$, therefore allowing us to eliminate the summation over $\vect{G}'$ in the last equation above. Note that in the penultimate step, we have made use of the relation $\vec{\mc{U}}_{\vect{k}+\vect{G}_1} (\vect{G}_2)$\,$=$\,$\vec{\mc{U}}_{\vect{k}} (\vect{G}_1+\vect{G}_2)$, which we can always enforce; this is equivalent to asserting $\vec{\Psi}^{}_{\vect{k}+\vect{G}} (\vect{r}) = \vec{\Psi}^{}_{\vect{k}} (\vect{r}) \, \forall \, \vect{G}$. We finally have
\begin{alignat}{1}
\label{eq:rho}
{\rho}^\pdagger_{\vect{q}} = \frac{1}{N} \sum_{\vect{k}, \sigma, \xi}  c^\dagger_{\vect{k},\sigma,\xi}\,c^\pdagger_{\mathrm{FBZ}(\vect{k}+\vect{q}),\sigma,\xi} \, F^{}_{\vect{k},\vect{q},\xi},
\end{alignat}
where we have introduced the form factors 
\begin{alignat}{1}
\label{eq:FF}
F^{}_{\vect{k},\vect{q},\xi} &\equiv \sum_{\vect{G}}\, \vec{\mc{U}}^\dagger_{\vect{k},\xi} (\vect{G})\,\, \vec{\mc{U}}^\pdagger_{\vect{k} + \vect{q},\xi} (\vect{G}),
\\
&= \sum_{\vect{G}}\, \vec{\mc{U}}^\dagger_{\vect{k},\xi} (\vect{G})\,\, \vec{\mc{U}}^\pdagger_{\vect{k} + \mathrm{FBZ} (\vect{q}),\xi} (\vect{G}+\vect{G}_{\vect{q}}),  \label{FinalFormOfFormfactors}
\end{alignat}
with $\vect{G}_{\vect{q}} = \vect{q}-\mathrm{FBZ} (\vect{q})$, i.e., the reciprocal lattice vector which folds the wavevector $\vect{q}$ back into the first Brillouin zone. The last form in \equref{FinalFormOfFormfactors} shows most clearly, why the form factors are expected to decay with large $|\vect{q}|$. 

Finally, let us elucidate how the $\vec{\mc{U}}^\pdagger_{\vect{k}}$ entering the form factors are computed from the continuum model in \appref{sec:model}. For notational simplicity, let us recast both spin and valley blocks of Eq.~\eqref{eq_AB-AB} in the form of a single Hamiltonian $\widetilde{H}_{\textrm{AB-AB}} = h ( - i \nabla) + \mc{T} (\vect{r})$, where $h$, which is a matrix in $\alpha$-space, has no explicit spatial dependence whereas $\mc{T}$ encapsulates the interlayer coupling and is moir\'{e}-lattice periodic. Appealing to translational symmetry and decomposing the latter in a Fourier series as $\mc{T} (\vect{r}) = \sum_{\vect{G}} \mc{T}_{\vect{G}}\, \exp \,(i \vect{G}.\vect{r})$, it is straightforward to show that the Schr\"{o}dinger equation $\widetilde{H}_{\textrm{AB-AB}} \left \lvert \Psi_{n,\vect{k}} \right \rangle = E^{}_{n,\vect{k}} \left \lvert \Psi_{n,\vect{k}} \right \rangle$ translates to
\begin{alignat}{1}
\nonumber
h (\vect{k}+ \vect{G})\,\, \vec{\mc{U}}^{}_{n,\vect{k}} (\vect{G}) + \sum_{\vect{G}'}\mc{T}^{}_{\vect{G}-\vect{G}'}\,\, \vec{\mc{U}}^{}_{n,\vect{k}} (\vect{G}') = E^{}_{n,\vect{k}}\,\, \vec{\mc{U}}^{}_{n,\vect{k}} (\vect{G})
\end{alignat}
$\forall \,\vect{k}\,\in\,$FBZ. This provides a prescription to read off $\vec{\mc{U}}^\pdagger_{\vect{k}}$ from the eigenvectors obtained upon diagonalizing the Hamiltonian $H_{\textrm{AB-AB}}$ in Eq.~\eqref{eq_AB-AB}. 

On a side note, we remark that the density, $\hat{\varrho}^{\pdagger}_{\vect{q},l,s}$, in a particular layer $\ell = (l,s)$, rather than the total density, can be projected in a similar fashion, $\hat{\varrho}^{\pdagger}_{\vect{q},l,s} \rightarrow {\varrho}^{\pdagger}_{\vect{q},l,s}$ as
\begin{alignat}{1}
\label{eq:rhol}
{\varrho}^{\pdagger}_{\vect{q},l,s} &= \frac{1}{N} \sum_{\vect{k}, \sigma, \xi}  c^\dagger_{\vect{k},\sigma,\xi}\,c^\pdagger_{\mathrm{FBZ}(\vect{k}+\vect{q}),\sigma,\xi} \, F^{l,s}_{\vect{k},\vect{q},\xi}
\end{alignat}
with
\begin{alignat}{1}
F^{l,s}_{\vect{k},\vect{q},\xi} &\equiv \sum_{\vect{G}}\, \vec{\mc{U}}^\dagger_{\vect{k},\xi} (\vect{G})\,P^{l,s}\,\, \vec{\mc{U}}^\pdagger_{\vect{k} + \vect{q},\xi} (\vect{G}),
\end{alignat}
where $P^{l,s}$ is a square diagonal matrix that projects on to layer $(l,s)$. By definition, it necessarily holds that $F_{\vect{k},\vect{q},\xi}$\,$=$\,$ \sum_{l,s} F^{l,s}_{\vect{k},\vect{q},\xi}$\,$\equiv$\,$\sum_{\ell}F^{\ell}_{\vect{k},\vect{q},\xi}$.\\

\section{Linearized mean-field gap equation}
\label{app:mf}
Having established the procedure for the projection of the density operators to the low-energy subspace formed by the single conduction band (per spin and valley), we are now well-positioned to construct a mean-field theory of superconductivity in TDBG. We start from the density-density interaction given by Eq.~\eqref{eq:Hint} and keep $V_{l,l'}^{s,s'}(\vect{q})$ general for now. We also emphasize that $\vect{q}$ in Eq.~\eqref{eq:Hint} extends over \textit{all} two-dimensional momenta but $V_{l,l'}^{s,s'}(\vect{q})$ need not be periodic in the moir\'{e} Brillouin zone. 

\begin{widetext}
Using the expression for the densities from \equref{eq:rhol} and the same multi-index notation, $\ell =(l,s)$ introduced in \appref{app:model}, we find 
\begin{equation}
H_{\mathrm{int}} = \sum_{\vect{k},\vect{k}',\vect{q}} V^{\ell,\ell'}_{\vect{q}} c^\dagger_{\vect{k},\sigma,\xi}\,c^\pdagger_{\mathrm{FBZ}(\vect{k}+\vect{q}),\sigma,\xi} c^\dagger_{\vect{k}',\sigma',\xi'}\,c^\pdagger_{\mathrm{FBZ}(\vect{k}'-\vect{q}),\sigma',\xi'} F^{\ell}_{\vect{k},\vect{q},\xi} F^{\ell'}_{\vect{k}',-\vect{q},\xi'},
\label{GeneralFormOfInteraction}\end{equation}
with $\vect{k}, \vect{k}'\in\,$FBZ. All repeated internal indices ($\sigma$, $\sigma'$, $\xi$, $\xi'$, $\ell$, $\ell'$) are summed over, and we adopt this convention for the remainder of the discussion as well. Furthermore, we omit the ``FBZ'' hereafter and implicitly assume that the momentum arguments of the field operators are folded back into the first Brillouin zone.  Retaining only terms in the homogeneous intervalley Cooper channel ($\vect{k}$\,$=$\,$-\vect{k}'$, $\xi$\,$=$\,$-\xi'$), $H_{\mathrm{int}} $ reduces to
\begin{alignat}{1}
H_{\mathrm{int}}^{\textsc{cc}}= -\sum_{\vect{k},\vect{q}} V^{\ell,\ell'}_{\vect{q}}\, F^{\ell}_{\vect{k},\vect{q},\xi}\, F^{\ell'}_{-\vect{k},-\vect{q},-\xi}\, c^\dagger_{\vect{k},\sigma,\xi}\, c^\dagger_{-\vect{k},\sigma',-\xi}\, c^\pdagger_{\vect{k}+\vect{q},\sigma,\xi}\, c^\pdagger_{-\vect{k}-\vect{q},\sigma',-\xi}.
\end{alignat}
The form factors in this channel can be simplified further: time-reversal symmetry dictates $F^{\ell}_{\vect{k},\vect{q},\xi}$\,$=$\,$(F^{\ell}_{-\vect{k},-\vect{q},-\xi})^*$, wherefore 
\begin{alignat}{1}\label{eq:Hint2}
H_{\mathrm{int}}^{\textsc{cc}} = \sum_{\vect{k},\vect{q}} \mc{C}^{}_{\vect{k},\vect{q}}\, c^\dagger_{\vect{k},\sigma,+}  c^\dagger_{-\vect{k},\sigma',-}  c^\pdagger_{\vect{k}+\vect{q},\sigma,+}  c^\pdagger_{-\vect{k}-\vect{q},\sigma',-}, \qquad \mc{C}^{}_{\vect{k},\vect{q}} &\equiv -\,\left( V^{\ell \ell'}_{\vect{q}} + V^{\ell \ell'}_{-\vect{q}} \right)\, F^{\ell}_{\vect{k},\vect{q},+} \left(F^{\ell'}_{\vect{k},\vect{q},+}\right)^* .
\end{alignat}
\end{widetext}
An advantage of Eq.~\eqref{eq:Hint2} is that it now suffices to compute the wavefunctions for a single valley, say, $\xi$\,$=$\,$+$, only. 
For the electron-phonon coupling discussed in the main text, one finds
\begin{equation}
    \mc{C}^{}_{\vect{k},\vect{q}} = \delta_{l,l'} \frac{D^2}{\bar{m}\, v^2_{ph}} \sum_{n} \frac{\omega^2_{\vect{q}}}{\omega^2_{\vect{q}}+ \nu_n^2} \sum_l \bigg|\sum_s F^{l,s}_{\vect{k},\vect{q},+}\bigg|^2,
\end{equation}
whereas for the screened Coulomb interaction, $V^{}_{\textsc{rpa}} (\vect{q})$,
\begin{equation}
    \mc{C}^{}_{\vect{k},\vect{q}} =  -(V^{}_{\textsc{rpa}} (\vect{q}) + V^{}_{\textsc{rpa}} (-\vect{q})) \left| F^{}_{\vect{k},\vect{q},+} \right|^2.
\end{equation}

The mean-field decoupling proceeds by defining, as usual, the expectation value
\begin{equation}
\label{eq:D1}
\left(\Delta^{\prime}_{\vect{k}}\right)_{\sigma, \sigma'} \equiv \big\langle c^\pdagger_{\vect{k},\sigma,+} c^\pdagger_{-\vect{k},\sigma',-} \big \rangle,
\end{equation}
whereupon we get (neglecting a constant piece)
\begin{equation}
H^{\textsc{mf}}_{\mathrm{int}} = \sum_{\vect{k},\vect{q} } \mc{C}^{}_{\vect{k},\vect{q}} \,c^\dagger_{\vect{k},\sigma,+} c^\dagger_{-\vect{k},\sigma',-} \left(\Delta^{\prime}_{\vect{k}+\vect{q}}\right)_{\sigma, \sigma'} + \mathrm{H.c.}\,.
\end{equation}
Compactly expressing the $\vect{q}$-sum as 
\begin{alignat}{1}
\label{eq:cons}
\left(\hat{\Delta}^{\pdagger}_{\vect{k}}\right)^{}_{\sigma, \sigma'} \equiv \sum_{\vect{q}} \mc{C}^{}_{\vect{k},\vect{q}} \left(\Delta^{\prime}_{\vect{k}+\vect{q}}\right)_{\sigma, \sigma'}, 
\end{alignat}
the mean-field interaction is consolidated into
\begin{equation}
H^{\textsc{mf}}_{\mathrm{int}} = \sum_{\vect{k}}^{\mathrm{FBZ}}c^\dagger_{\vect{k},\sigma,+}  \left(\hat{\Delta}^{}_{\vect{k}}\right)^{}_{\sigma, \sigma'}  c^\dagger_{-\vect{k},\sigma',-}  + \mathrm{H.c.}\,.
\end{equation}
In terms of the Nambu spinor $\psi_{\vect{k},\sigma}^\mathrm{T}$\,$=\,$\,$(c^{}_{\vect{k},\sigma,+},\, c^{\dagger}_{-\vect{k},\sigma,-} )^\mathrm{T}$ \cite{PhysRev.117.648}, the \textit{full} Hamiltonian, including both hopping and pairing, can be kneaded into the BdG form of Eq.~\eqref{BdGFormofHam},
taking advantage of the fact that the dispersions in the two valleys are related as $E^{}_{\vect{k},+}$\,$=$\,$E^{}_{-\vect{k}, -}$ as a consequence of time-reversal symmetry. All the microscopic details of TDBG are encoded in this mean-field Hamiltonian through the band energies and the form factors.

For self-consistency, Eq.~\eqref{eq:cons} is required to hold, where the expectation value on the right-hand side [see \equref{eq:D1}] is calculated within $H^{\textsc{mf}}$. We have
\begin{alignat}{1}
\nonumber
\left \langle c^\pdagger_{\vect{k},\sigma,+} c^\pdagger_{-\vect{k},\sigma',-} \right \rangle = - T \sum_{\omega_n} \left[\mc{G}^{}_\mathrm{BdG}(\vect{k},\omega_n) \right]_{\substack{1,2\\ \sigma, \sigma'}};
\end{alignat}
the numbers $1,2$ denote the indices in Nambu space, while $\sigma$ and $\sigma'$ are spin indices, and  the BdG Green's function is given by $\mc{G}^{}_\mathrm{BdG}(\vect{k},\omega_n)=(i \omega_n - h^{}_\mathrm{BdG} (\vect{k}))^{-1}$. We expand in the superconducting order parameter,
$\mc{G}^{}_\mathrm{BdG} = \left[\mc{G}^{-1}_0 - \Sigma^{sc}\right]^{-1} \sim \mc{G}^{}_0 +  \mc{G}^{}_0 \Sigma^{sc} \mc{G}^{}_0$, where
\begin{alignat}{1}
\mc{G}^{-1}_0 &= \left[i \omega_n - 
\begin{pmatrix} 
E^{}_{\vect{k},+}  &  0 \\
0  & -E^{}_{\vect{k},+} 
\end{pmatrix}
 \right] \delta_{\sigma, \sigma'} ,\\
\Sigma^{sc} &= \begin{bmatrix} 
0 &  \left(\hat{\Delta}^{\pdagger}_{\vect{k}}\right)^{}_{\sigma, \sigma'}  \\
 \left(\hat{\Delta}^{\dagger}_{\vect{k}}\right)^{}_{\sigma, \sigma'}  & 0
\end{bmatrix},
\end{alignat}
and get, to linear order in $\hat{\Delta}$:
\begin{alignat}{1}
\nonumber \big\langle c^\pdagger_{\vect{k},\sigma,+} c^\pdagger_{-\vect{k},\sigma',-}\big\rangle &= \sum_{\omega_n}\frac{-T}{(i \omega_n - E^{}_{\vect{k},+})(i \omega_n + E^{}_{\vect{k},+})} \left(\hat{\Delta}^{}_{\vect{k}}\right)^{}_{\sigma, \sigma'}\\
&= \frac{1}{2\, E^{}_{\vect{k},+}}\tanh \left( \frac{E^{}_{\vect{k},+}}{2 T}\right) \left(\hat{\Delta}^{}_{\vect{k}}\right)^{}_{\sigma, \sigma'}.
\label{eq:sc_mat}
\end{alignat}
Equation \eqref{eq:cons} now becomes
\begin{alignat}{1}
\label{eq:sc2}
 \left(\hat{\Delta}^{}_{\vect{k}}\right)^{}_{\sigma, \sigma'} = \sum_{\vect{k}'}  \frac{\mc{V}_{\vect{k},\vect{k}'}}{2\, E^{}_{\vect{k},+}}\tanh \left( \frac{E^{}_{\vect{k},+}}{2 T}\right) \left(\hat{\Delta}^{}_{\vect{k}'}\right)^{}_{\sigma, \sigma'},
\end{alignat}
defining
\begin{equation}
\label{eq:v_ref}
\mc{V}_{\vect{k},\vect{k}'} \equiv \mc{C}^{}_{\vect{k},\vect{k}'-\vect{k}}.
\end{equation}
It is easy to observe that
\begin{align}
    \mc{V}_{\vect{k},\vect{k}'} &= \mc{V}^*_{\vect{k}',\vect{k}}= \mc{V}_{\vect{k} +\vect{G},\vect{k}'+\vect{G}}. \label{VGTransl}
\end{align}
In \equref{eq:sc2}, we now see explicitly that singlet and triplet are degenerate as expected due to the presence of the SO(4)$\,\simeq\,$SU(2)$\times$SU(2) symmetry of independent spin rotations in each valley: inserting either a singlet, $(\hat{\Delta}_{\vect{k}})_{\sigma, \sigma'} $\,$\equiv$\,$ \Delta_{\vect{k}}\, (i \sigma_y)_{\sigma, \sigma'}$, or a triplet ansatz, $(\hat{\Delta}_{\vect{k}})_{\sigma, \sigma'} $\,$\equiv$\,$ \Delta_{\vect{k}}\, (i \sigma_y \vect{\sigma}\cdot \vect{d})_{\sigma, \sigma'}$, we obtain the same eigenvalue equation \eqref{eq:gapeq}.


Obviously, due to \equref{VGTransl}, $\mc{V}$ is Hermitian (it is, in our case, real and symmetric) which guarantees that the kernel $\mathcal{M}$ only has real eigenvalues. In general, the sum over $\vect{k}'$ in \equref{eq:gapeq1} involves arbitrarily large momenta. However, due to the decay of the form factors with large momentum transfer $\vect{q} = \vect{k}'-\vect{k}$ in \equref{eq:Hint2}, we restrict the sum to $\vect{k}'$ in the FBZ.

At sufficiently large $T$, $\mc{M}^{}_{\vect{k}, \vect{k}'} (T)$\,$\sim$\,$1/T$ and Eq.~\eqref{eq:gapeq1} does not have a solution. If, below a critical temperature $T_c$, a solution exists for the order parameter, its structure in momentum space is, of course, interaction-dependent: when $V_{\vect{k}'-\vect{k}} > 0$ $(< 0)$, i.e., the interaction is repulsive (attractive), $\Delta_{\vect{k}}$ and  $\Delta_{\vect{k}'}$ are favored to have the opposite (same) sign. 

\section{Selection of singlet or triplet pairing}
\label{app:singlettriplet}
In this appendix, we discuss in more detail the possibility that the dominant $\text{SU}(2)_+ \times \text{SU}(2)_-$-preserving pairing interactions are provided by phonons while those breaking this enhanced spin symmetry---down to the usual total SU(2) spin symmetry---are dominated by purely electronic physics. Specifically, we show that triplet (singlet) pairing is generically favored if the latter class of interactions are dominated by time-reversal odd (even) particle-hole fluctuations. 

To this end, let us assume that, at the energy scales relevant to superconductivity, there is a set of collective electronic modes, $\phi_{\vect{q}}^{j}$, $j$\,$=$\,$1,2,\dots N_b$, that dominates the part of the pairing mechanism that breaks the $\text{SU}(2)_+ \times \text{SU}(2)_-$ symmetry (and, thus, determines whether singlet or triplet will win). We do not want to make any assumptions about its microscopic form and, thus, take the general form of the coupling to the electrons given in \equref{GeneralFermionBosonCoupling} of the main text.
As we will see below, it is sufficient to specify the behavior of $\phi_{\vect{q}}^{j}$ under time-reversal symmetry: we will focus on all of these modes being time-reversal even, $t_\phi=+$, or odd, $t_\phi=-$, corresponding to
\begin{equation}
    \hat{\Theta}\, \phi^\pdagger_{\vect{q}}\,\hat{\Theta}^\dagger = t_\phi\,\phi_{-\vect{q}}, \,\, \Theta\, \lambda(-\vect{k},-\vect{k}')\,\Theta^\dagger = t_\phi\, \lambda(\vect{k},\vect{k}'), \label{TimeReversalConstr}
\end{equation}
where $\hat{\Theta}$ and $\Theta$ are the anti-unitary time-reversal operators in Fock and single-particle space, respectively. 

For the remainder of this section, we use the field-integral description and denote the associated field operators of the electrons and collective bosonic modes by the same symbols, $c_{k,\alpha}$ and $\phi_{q}^{j}$, employing the combined Matsubara-frequency and momentum notation, $k$\,$\equiv$\,$(i\omega_n,\vect{k})$ and $q$\,$\equiv$\,$(i\nu_n,\vect{q})$; here $\omega_n$ and $\nu_n$ are fermionic and bosonic Matsubara frequencies, respectively. The effective low-energy dynamics of the collective bosonic modes is described by
\begin{equation}
    \mathcal{S}_{\phi} = \frac{1}{2}\int_q \phi_{q}^j \left[\chi^{-1}(i\nu^{}_n,\vect{q})\right]_{j,j'} \phi_{-q}^{j'},
\end{equation}
where $\int_q \dots \equiv T\sum_{\nu_n} \sum_{\vect{q}} \dots$ and $\chi^{-1}(i\nu_n,\vect{q})$ is the (full) susceptibility in the corresponding particle-hole channels. It can be shown that \cite{GeneralProof} 
\begin{align}
    \chi\,(i\nu^{}_n,\vect{q}) &= \chi^T(-i\nu^{}_n,-\vect{q}), \label{ChiSymmetric} \\
    \chi\,(i\nu^{}_n,\vect{q}) &= \chi^\dagger(i\nu^{}_n,\vect{q}). \label{ChiHermitian}
\end{align}
While the first property simply follows from $\chi$ being a correlator of twice the same real bosonic field, the second equation is related to time-reversal symmetry and Hermiticity. Apart from $\mathcal{S}_{\phi}$, the total action $\mathcal{S} = \mathcal{S}_0 + \mathcal{S}_\phi + \mathcal{S}_{\text{int}}+\mathcal{S}_{c\phi}$ contains the free electronic theory,
\begin{equation}
    \mathcal{S}_0 = \int_k c^\dagger_{k,\alpha} \left[-i\omega^{}_n + h^{}_{\alpha\beta}(\vect{k})\right] c^\pdagger_{k,\beta},
\end{equation}
the fermion-boson coupling, $\mathcal{S}_{c\phi}$, analogous to \equref{GeneralFermionBosonCoupling}, and the contribution $\mathcal{S}_{\text{int}}$ describing all other interactions preserving the $\text{SU}(2)_+ \times \text{SU}(2)_-$ symmetry [such as those in \equref{GeneralFormOfInteraction}].

To study superconductivity, we use the following low-energy description. We diagonalize the free Hamiltonian,
\begin{equation}
h(\vect{k})\, \psi^{}_{\vect{k}\xi\sigma} = E^{}_{\vect{k}\xi} \,\psi^{}_{\vect{k}\xi\sigma}, \label{DiagonalizeHamiltonian}
\end{equation}
where we have already taken into account that spin-orbit coupling can be neglected (spin $\sigma$ is a good quantum number) and further ignored any intervalley mixing on the level of the noninteracting Hamiltonian $h(\vect{k})$. These mixing terms are very small leading to the approximate valley-charge-conservation symmetry, U(1)$_v$, which allows us to focus on intervalley pairing \cite{2019arXiv190603258S}. Breaking U(1)$_v$ weakly (as is the case for the real system) will generate a small admixture of intravalley pairing, which is, however, not of interest to our discussion here. In \equref{DiagonalizeHamiltonian} and the following, we concentrate on the eigenstates that give rise to Fermi surfaces and project the electronic fields according to $c_{k\alpha} \rightarrow \sum_{\xi,\sigma} (\psi_{\vect{k}\xi\sigma})_\alpha f_{k\xi\sigma}$. 

\begin{widetext}
Integrating out the bosonic modes $\phi_{q}^j$, we obtain an effective action $\mathcal{S}^f = \mathcal{S}_0^f + \mathcal{S}_{\text{int}}^f$ with free contribution $\mathcal{S}^f_0 = \int_k f^\dagger_{k\xi\sigma} (-i\omega_n + E_{\vect{k}\xi}) f^\pdagger_{k\xi\sigma}$ and electron-electron interactions described by $\mathcal{S}_{\text{int}}^f$.  The relevant Cooper channel of $\mathcal{S}_{\text{int}}^f$ can be written as
\begin{equation}
    \mathcal{S}_{\text{cc}}^f = \int_{k}\int_{k'} \mathcal{V}^{\sigma_1\sigma_2}_{\sigma_3\sigma_4}(k',\xi';k,\xi)\, f^\dagger_{k'\xi'\sigma_1}f^\dagger_{-k'-\xi'\sigma_2}f^\pdagger_{-k-\xi\sigma_3}f^\pdagger_{k\xi\sigma_4}, \label{FullInteractionCooperCh}
\end{equation}
where $\mathcal{V}^{\sigma_1\sigma_2}_{\sigma_3\sigma_4}(k',\xi';k,\xi) = \mathcal{W}^{\sigma_1\sigma_2}_{\sigma_3\sigma_4}(k',\xi';k,\xi) + t_\phi\,\Delta \mathcal{V}^{\sigma_1\sigma_2}_{\sigma_3\sigma_4}(k',\xi';k,\xi)$ is the sum of the $\text{SU}(2)_+ \times \text{SU}(2)_-$-preserving interactions ($\mathcal{W}$) and the additional contribution ($\Delta \mathcal{V}$) from the collective bosonic modes. Making use of \equref{TimeReversalConstr} and the fact that time-reversal symmetry implies $\Theta\, \psi_{\vect{k}\xi\sigma} = s_\sigma \psi_{-\vect{k}-\xi\bar{\sigma}}$, where $s_{\uparrow} =+$, $s_{\downarrow} = -$ and $\bar{\uparrow} = \downarrow$, $\bar{\downarrow} = \uparrow$, the latter can be rewritten as
\begin{equation}
    \Delta \mathcal{V}^{\sigma_1\sigma_2}_{\sigma_3\sigma_4}(k',\xi';k,\xi) = -\frac{1}{2} \sigma^{}_2\,\sigma^{}_3\, \Lambda^{j*}_{\xi'\overline{\sigma_2},\xi\overline{\sigma_3}}(\vect{k}',\vect{k})\,[\chi(k'-k)]^{}_{j,j'}\,\Lambda^{j'}_{\xi'\sigma_1,\xi\sigma_4}(\vect{k}',\vect{k}),  \label{DeltaVContr}
\end{equation}
introducing the shorthand $\Lambda^{j}_{\xi\sigma,\xi'\sigma'}(\vect{k},\vect{k}') = \psi^\dagger_{\vect{k}\xi\sigma}\,\lambda^{j}(\vect{k},\vect{k}')\,\psi^\pdagger_{\vect{k}'\xi'\sigma'}$.
\end{widetext}
Next, we decouple, as usual, the interaction in \equref{FullInteractionCooperCh} in the Cooper channel with the help of the Hubbard-Stratonovich fields $\Delta_{\sigma\sigma'}^\xi(k)$ and $\overline{\Delta}_{\sigma\sigma'}^\xi(k)$. To linear order in these fields, the saddle-point equation with respect to $\overline{\Delta}_{\sigma\sigma'}^\xi(k)$ can be written as
\begin{equation}
    \Delta_{\sigma\sigma'}^\xi(k) = -\int_{k'} \frac{\mathcal{V}^{\sigma\sigma'}_{\widetilde{\sigma}\widetilde{\sigma}'}(k,\xi;k',\xi')}{\omega_{n'}^2+ E^2_{\vect{k}'\xi'}} \Delta_{\widetilde{\sigma}'\widetilde{\sigma}}^{\xi'}(k'). \label{SaddlePointEq}
\end{equation}
As the system has SU(2) spin-rotation symmetry and the enhanced $\text{SU}(2)_+ \times \text{SU}(2)_-$ symmetry is broken, we know that we can discuss spin singlet and triplet separately. Beginning with the former, we write $\Delta_{\sigma\sigma'}^\xi(k) = s_\sigma \delta_{\bar{\sigma},\sigma'} \Delta^{\xi}_k$, with $\Delta^{\xi}_k=\Delta^{-\xi}_{-k}$. The saddle-point equation (\ref{SaddlePointEq}) leads to
\begin{equation}
    \widetilde{\Delta}^{\xi}_k = \int_{k'}\frac{F^0_{\xi k,\xi' k'}+t_\phi\,F^s_{\xi k,\xi' k'}}{(\omega_{n}^2+ E^2_{\vect{k}\xi})^{\frac{1}{2}}(\omega_{n'}^2+ E^2_{\vect{k}'\xi'})^{\frac{1}{2}}} \widetilde{\Delta}^{\xi'}_{k'},
\end{equation}
where we introduced $\widetilde{\Delta}^\xi_{k} = \Delta_{k}^\xi/(\omega_{n}^2+ E^2_{\vect{k}\xi})^{1/2}$ (to make the kernel Hermitian). $F^0_{\xi k,\xi' k'}$ is related to the $\text{SU}(2)_+ \times \text{SU}(2)_-$-symmetric $\mathcal{W}$, while the contribution of the collective electronic modes, i.e., $\Delta\mathcal{V}$ in \equref{DeltaVContr}, is contained in
\begin{equation}
    F^s_{\xi k,\xi' k'} = \frac{1}{2} \sum_{\sigma,\sigma'} \Lambda^{j*}_{\xi\sigma,\xi'\sigma'}(\vect{k},\vect{k}')\,[\chi(k-k')]^{}_{j,j'}\,\Lambda^{j'}_{\xi\sigma,\xi'\sigma'}(\vect{k},\vect{k}').
\end{equation}
Note that, due to \equref{ChiHermitian}, $\chi(q)$ has real eigenvalues, all of which must be positive as required by stability; hence, we find $F^s_{\xi k,\xi k'}$\,$>$\,$0$. Furthermore, $F^s$ is symmetric, $F^s_{\xi k,\xi k'}$\,$=$\,$F^s_{\xi' k',\xi k}$, as follows from \equref{ChiSymmetric}. Like previously, we view the saddle-point equation as a matrix equation, $\widetilde{\Delta}\, \lambda(T)$\,$=$\,$\mathcal{M}(T)\, \widetilde{\Delta}$ with a Hermitian matrix $\mathcal{M}(T)$: upon decreasing temperature, the largest eigenvalue $\lambda(T)$ increases and reaches $1$ at $T_c$. Denoting the eigenvalue when setting $F^s_{\xi k,\xi k'}$\,$\rightarrow$\,$0$ by $\lambda_0(T)$ and treating $F^s_{\xi k,\xi' k'}$ as a perturbation yields $\lambda(T) \sim \lambda_0(T) + t_\phi \delta\lambda_s(T)$ at leading order, where
\begin{equation}
    \delta\lambda_s(T) = \int_{k}\int_{k'} \frac{(\widetilde{\Delta}^\xi_{k})^*F^s_{\xi k,\xi' k'}\widetilde{\Delta}^{\xi'}_{k'}}{(\omega_{n}^2+ E^2_{\vect{k}\xi})^{\frac{1}{2}}(\omega_{n'}^2+ E^2_{\vect{k}'\xi'})^{\frac{1}{2}}} .\label{deltalambdaSingl}
\end{equation}
Similarly, for triplet, with, say, $\Delta_{\sigma\sigma'}^\xi(k) = \delta_{\bar{\sigma},\sigma'}  \Delta_k^\xi$, $\Delta_k^\xi = -\Delta_{-k}^{-\xi}$, we find the same equations as above, only with $F^s_{k,k'}$ replaced by $F^t_{k,k'} =$
\begin{alignat*}{1}
    \frac{1}{2} \sum_{\sigma,\sigma'} s^{}_\sigma s^{}_{\sigma'}\Lambda^{j*}_{\xi\sigma,\xi'\sigma'}(\vect{k},\vect{k}')\,[\chi(k-k')]_{j,j'}\,\Lambda^{j'}_{\xi\sigma,\xi'\sigma'}(\vect{k},\vect{k}').
\end{alignat*}
Most importantly, the eigenvalue will behave as $\lambda(T) \sim \lambda_0(T) + t_\phi \delta\lambda_t(T)$, where $\delta\lambda_t(T)$ is given by \equref{deltalambdaSingl} but with $F^s_{k,k'}$\,$\rightarrow$\,$F^t_{k,k'}$.
Since $\widetilde{\Delta}^+_k$\,$>$\,$0$ (without loss of generality) if electron-phonon coupling dominates $F^0$, as proven in \refcite{GeneralProof} and can, for example, be seen in Fig.~\ref{fig:Phonon}\,(d), we conclude that $\delta \lambda_s(T) > \delta \lambda_t(T)$. Therefore, the eigenvalue of singlet (triplet) will be larger and, hence, singlet (triplet) will dominate for $t_\phi = +$ ($t_\phi = -$).


\end{appendix}

\bibliography{Refs}

\end{document}